\documentclass{aa}

\usepackage{graphicx}
\usepackage{epsfig}

\newcommand{\pl}{\partial}
\renewcommand{\d}{{\rm d}}
\newcommand{\inta}{\int_{-i\infty}^{+i\infty}}
\newcommand{\beq}{\begin{equation}}
\newcommand{\eeq}{\end{equation}}
\newcommand{\beqa}{\begin{eqnarray}}
\newcommand{\eeqa}{\end{eqnarray}}
\newcommand{\bea}{\begin{array}}
\newcommand{\ea}{\end{array}}
\newcommand{\bx}{{\bf x}}

\newcommand{\cG}{{\cal G}}

\newcommand{\rhob}{\overline{\rho}}
\newcommand{\bk}{{\bf k}}
\newcommand{\lag}{\langle}
\newcommand{\rag}{\rangle}

\newcommand{\Om}{\Omega_{\rm m}}
\newcommand{\Ol}{\Omega_{\Lambda}}

\newcommand{\dR}{\delta_{R}}

\newcommand{\dL}{\delta_L}
\newcommand{\DL}{\Delta_L}
\newcommand{\DLo}{\Delta_{L0}}
\newcommand{\cN}{{\cal N}}

\newcommand{\cF}{{\cal F}}

\newcommand{\Sa}{S_{\alpha}}
\newcommand{\DW}{\Delta_W}
\newcommand{\np}{n_{\phi}}
\newcommand{\Dp}{\Delta_{\phi}}
\newcommand{\sigp}{\sigma_{\phi}}
\newcommand{\Det}{{\rm Det}}
\newcommand{\Tr}{{\rm Tr}}
\newcommand{\Dla}{\Delta_{\lambda}}
\newcommand{\cD}{{\cal D}}
\newcommand{\psib}{\overline{\psi}}

\renewcommand{\Im}{{\rm Im}}

\newcommand{\cP}{{\cal P}}

\begin{document}

 

\title{Dynamics of gravitational clustering III. The quasi-linear regime for some non-Gaussian initial conditions.}   
\author{P. Valageas}  
\institute{Service de Physique Th\'eorique, CEN Saclay, 91191 Gif-sur-Yvette, France} 
\date{Received / Accepted }

\abstract{
Using a non-perturbative method developed in a previous work (\cite{paper2}), we derive the probability distribution $\cP(\dR)$ of the density contrast within spherical cells in the quasi-linear regime for some non-Gaussian initial conditions. We describe three such models. The first one is a straightforward generalization of the Gaussian scenario. It can be seen as a phenomenological description of a density field where the tails of the linear density contrast distribution would be of the form $\cP_L(\dL) \sim e^{-|\dL|^{-\alpha}}$, where $\alpha$ is no longer restricted to 2 (as in the Gaussian case). We derive exact results for $\cP(\dR)$ in the quasi-linear limit. The second model is a physically motivated isocurvature CDM scenario. Our approach needs to be adapted to this specific case and in order to get convenient analytical results we introduce a simple approximation (which is not related to the gravitational dynamics but to the initial conditions). Then, we find a good agreement with the available results from numerical simulations for the pdf of the linear density contrast for $\delta_{L,R} \ga 0$. We can expect a similar accuracy for the non-linear pdf $\cP(\dR)$. Finally, the third model corresponds to the small deviations from Gaussianity which arise in standard slow-roll inflation. We obtain exact results for the pdf of the density field in the quasi-linear limit, to first-order over the primordial deviations from Gaussianity.
}

\titlerunning{The quasi-linear regime for some non-Gaussian initial conditions.}

\maketitle

\section{Introduction}

In usual cosmological scenarios, large-scale structures in the universe are generated by the growth of small primordial density perturbations, through gravitational instability. At large scales or at early times one can use a perturbative approach to obtain the first few moments of the evolved density field. Moreover, as described in \cite{Ber1} and \cite{Ber2} one can actually sum up the perturbative series at leading order in the limit $\sigma \rightarrow 0$ (where $\sigma$ is the rms density fluctuation) to obtain all order cumulants. This yields the probability distribution function (pdf) $\cP(\dR)$ of the density contrast within spherical cells. However, this method only applies to Gaussian initial conditions. Moreover, it may miss some non-perturbative effects.

In a previous paper (\cite{paper2}) we developed a non-perturbative method to derive the pdf $\cP(\dR)$ in this quasi-linear regime. It is based on a steepest-descent approximation which yields asymptotically exact results in this limit. This allowed us to justify the results obtained by perturbative means and to correct some misconceptions related to non-perturbative effects. Another advantage of our approach is that in principle it can also be applied to non-Gaussian primordial density fluctuations. However, the determination of the relevant saddle-points needed in order to get simple analytic results may then be more difficult. In particular, the method may need to be adapted to specific cases.

Thus, in this article we show in details how to obtain the pdf $\cP(\dR)$ for three specific non-Gaussian models, in the quasi-linear regime. Indeed, although observations are consistent with Gaussian initial conditions so far, it is of interest to investigate a broader class of models until any non-Gaussianity is definitely ruled out. Moreover, there exist some reasonable physical scenarios (though somewhat more contrived than the standard CDM model) which give rise to non-Gaussian primordial fluctuations. Finally, even standard inflationary scenarios lead to small deviations from Gaussianity for the primordial density fluctuations.

This article is organized as follows. First, in Sect.\ref{Generating functions} we recall the path-integral formulation which allows one to write an explicit expression for the pdf $\cP(\dR)$ in terms of initial conditions. Then, in Sect.\ref{Generalization of the Gaussian weight} we apply our method to a non-Gaussian model which is a straightforward generalization of the Gaussian case and could be used as a phenomenological tool to reproduce observations in case some departure from Gaussianity would be measured (e.g., in the CMB data). Then, in Sect.\ref{IsoCDM model} we investigate a physically motivated model which describes an isocurvature cold dark matter scenario (\cite{Peebles3}). In particular, we obtain a good agreement with the available results from numerical simulations for the pdf of the linearly evolved density field. Finally, in Sect.\ref{Standard slow-roll inflation} we consider the small primordial deviations from Gaussianity which arise in standard slow-roll inflation.

\section{Generating functions. Gaussian initial conditions}
\label{Generating functions}

In this article, we investigate the probability distribution function (pdf) $\cP(\dR)$ of the density contrast $\dR$ within a spherical cell of comoving radius $R$, volume $V$:
\beq
\dR = \int_V \frac{\d^3x}{V} \; \delta(\bx) .
\label{dR}
\eeq
Here $\delta(\bx)$ is the non-linear density contrast at the comoving coordinate $\bx$, at the time of interest. In this section, following the method developed in \cite{paper2}, we recall how one can express the pdf $\cP(\dR)$ as a function of the initial conditions through a path-integral formalism. In order to introduce our approach we first briefly consider the case of Gaussian initial conditions which is most familiar. In the next section, we extend our formalism to a non-Gaussian model which can be seen as the simplest generalization of the usual Gaussian case.

Rather than trying to directly evaluate the pdf itself, it is actually more convenient to study its Laplace transform $\psi(y)$ given by:
\beq
\psi(y) \equiv \lag e^{-y \dR} \rag \equiv \int_{-1}^{\infty} \d\dR \; e^{-y \dR} \; \cP(\dR)  .
\label{psi1}
\eeq
Here, the symbol $\lag .. \rag$ expresses the average over the initial conditions. Then, the last term in eq.(\ref{psi1}) can also be seen as the definition of the pdf $\cP(\dR)$. Indeed, the pdf can be recovered from $\psi(y)$ through the standard inverse Laplace transform:
\beq
\cP(\dR) = \inta \frac{\d y}{2\pi i} \; e^{y \dR} \; \psi(y) .
\label{P1}
\eeq
Moreover, the generating function $\psi(y)$ also yields the moments $\lag \dR^q \rag$ through:
\beq
\psi(y) = \sum_{q=0}^{\infty} \frac{(-y)^{q}}{q!} \; \lag \dR^q \rag .
\label{psiseries1}
\eeq

Then, we need to compute $\psi(y)$ as an average over the initial conditions, using the first equality in eq.(\ref{psi1}). To do so, one simply needs two pieces of information. First, we must know the functional $\dR[\dL(\bx)]$ which yields the exact non-linear density contrast $\dR$ over the cell $V$ which arises from the gravitational dynamics of the linear density field $\dL(\bx)$. Indeed, as shown for instance in \cite{paper1} the initial conditions can be defined by the linear density contrast $\dL(\bx)$ where we only keep the linear growing mode. This does not assume that the exact non-linear density field $\delta(\bx)$ can be written as a series expansion over powers of the linear field $\dL(\bx)$. In fact, such a series is only asymptotic (e.g., \cite{paper1}, \cite{paper5}). Hereafter, we note $\dL(\bx)$ the growing mode of the linear density contrast at the time of interest, where we compute $\cP(\dR)$ (i.e., to simplify notations we do not write explicitly the time dependence). Second, in order to perform the average in eq.(\ref{psi1}) we need the weight which is associated with all possible fields $\dL(\bx)$. In other words, we must specify the probability distribution of the random field $\dL(\bx)$. Thus, the first point expresses the physics of gravitational interactions while the second point describes the initial conditions of the system.

In the case of Gaussian initial conditions, the statistics of the random field $\dL(\bx)$ are fully defined by the two-point correlation:
\beq
\DL(\bx_1,\bx_2) \equiv \lag \dL(\bx_1) \dL(\bx_2) \rag .
\label{Dl1}
\eeq
The kernel $\DL$ is symmetric, homogeneous and isotropic: $\DL(\bx_1,\bx_2) = \DL(|\bx_1-\bx_2|)$. It is convenient to express $\DL$ in terms of the power-spectrum $P(k)$ of the linear density fluctuations. To do so, we define the Fourier transform of the density field as:
\beq
\left\{ \bea{l} {\displaystyle \delta(\bx) = \int \d\bk \; e^{i \bk.\bx} \; \delta(\bk) }  \\ {\displaystyle \delta(\bk) = \frac{1}{(2\pi)^3} \int \d\bx \; e^{-i \bk.\bx} \; \delta(\bx) } \ea \right.
\label{Four1}
\eeq
Next, we define the Fourier transform of the kernel $\DL$ by the property:
\beq
f_1 . \DL . f_2 = \int \d\bk_1 \d\bk_2 \; f_1(\bk_1)^{\ast} . \DL(\bk_1,\bk_2) . f_2(\bk_2)
\label{FourD}
\eeq
for any real fields $f_1$ and $f_2$, where we introduced the short-hand notation:
\beq
f_1 . \DL . f_2 \equiv \int \d\bx_1 \d\bx_2 \; f_1(\bx_1) . \DL(\bx_1,\bx_2) . f_2(\bx_2) .
\label{Dl0}
\eeq
Using eq.(\ref{Four1}) this implies:
\beq
\DL(\bk_1,\bk_2) = \int \d\bx_1 \d\bx_2 \; e^{i(\bk_2.\bx_2 - \bk_1.\bx_1)} \; \DL(\bx_1,\bx_2)
\label{KerFour1}
\eeq
which gives:
\beq
\DL(\bk_1,\bk_2) = (2\pi)^6 \; P(k_1) \; \delta_D(\bk_1 - \bk_2)
\label{Dl2}
\eeq
where we defined the power-spectrum $P(k)$ of the linear density contrast by:
\beq
\lag \dL(\bk_1) \dL(\bk_2) \rag \equiv P(k_1) \; \delta_D(\bk_1+\bk_2) .
\label{Pk1}
\eeq
Finally, the inverse $\DL^{-1}$ of the kernel $\DL$ is (see \cite{paper2}):
\beq
\DL^{-1}(\bk_1,\bk_2) = \frac{1}{P(k_1)} \; \delta_D(\bk_1-\bk_2)
\label{Dl3}
\eeq
which implies that $\DL^{-1}$ is positive definite since we have:
\beq
\dL . \DL^{-1} . \dL = \int \d\bk \; \frac{|\dL(\bk)|^2}{P(k)}
\label{Dl4}
\eeq
where we used $\dL(-\bk)=\dL(\bk)^{\ast}$ for real fields $\dL(\bx)$, see eq.(\ref{Four1}).

Thus, for Gaussian initial conditions the statistical properties of the random field $\dL(\bx)$ are determined by the kernel $\DL$ given in eq.(\ref{Dl2}) (or equivalently by $\DL^{-1}$ given in eq.(\ref{Dl3})). In particular, the average over the initial conditions in eq.(\ref{psi1}) can be written as the path-integral:
\beq
\psi(y) = \left( \Det\DL^{-1} \right)^{1/2} \int [\d\dL(\bx)] \; e^{-y \dR[\dL] -\frac{1}{2} \dL . \DL^{-1} . \dL} .
\label{psi2}
\eeq
This expression merely means that in order to compute the average $\lag e^{-y \dR} \rag$ we simply need to sum up the contributions of all possible linear density fields $\dL(\bx)$ to which we associate a Gaussian weight which is proportional to $e^{-(\dL.\DL^{-1}.\dL)/2}$. The normalization factor $(\Det\DL^{-1})^{1/2}$ ensures that $\psi(0)=1$, as implied by the definition (\ref{psi1}). Here $\Det\DL^{-1}$ is the determinant of the kernel $\DL^{-1}$. Thus, eq.(\ref{psi2}) yields an explicit expression for the Laplace transform $\psi(y)$ (though it may be difficult to obtain numerical results from such a path-integral). This provides in turn the pdf $\cP(\dR)$ through the inverse transform (\ref{P1}). In \cite{paper2} we showed how to get the generating function $\psi(y)$ from eq.(\ref{psi2}) in the limit $P(k) \rightarrow 0$ (i.e. for small rms density fluctuations) using a steepest-descent method. In the next section, we apply this method to a closely related non-Gaussian model.

\section{Generalization of the Gaussian weight}
\label{Generalization of the Gaussian weight}

In this section, we describe a non-Gaussian model which can be seen as the simplest generalization of the Gaussian scenario. This model is not derived from physical principles. It simply provides a phenomenological tool which can describe initial conditions such that the tails of the linear pdf $\cP_L(\dL)$ of the density field are of the form $\cP_L(\dL) \sim e^{-|\dL|^{-\alpha}}$, where $\alpha$ is no longer restricted to 2 (as in the Gaussian case). In addition, it allows us to show on a simple example the power of the method we developed in \cite{paper2} and how it can be applied to non-Gaussian initial conditions.

\subsection{A non-Gaussian model}
\label{The model}

In the case of Gaussian primordial density fluctuations, the average involved in the definition (\ref{psi1}) of the generating function $\psi(y)$ (i.e. the Laplace transform of the pdf $\cP(\dR)$) is simply given by the Gaussian weight $\exp(-\frac{1}{2} \dL . \DL^{-1} . \dL)$ in the path-integral (\ref{psi2}) which takes the average of $e^{-y \dR}$ over all possible initial states, as we recalled in the previous section. A straightforward generalization of the Gaussian case (\ref{psi2}) is to modify this weight. Thus, one can consider the case where the average over the field $\dL$ is given by:
\beq
\psi(y) = \cN \int [\d\dL(\bx)] \; e^{-y \dR - W[\dL]}
\label{psiNG1}
\eeq
with:
\beq
W[\dL] = \sum_{i=1}^{N} \lambda_i \; (\dL . \DW^{-1} . \dL)^{\alpha_i/2}
\label{Wd1}
\eeq
where $\alpha_1 < .. < \alpha_N$ and $\lambda_N>0$. Here $\cN$ is a normalization constant, so that $\psi(0)=1$, and we still take the kernel $\DW^{-1}$ to be of the form (\ref{Dl3}) hence $\DW^{-1}$ is positive definite. We also define a quantity $\sigma_W$ in terms of $\DW$ by:
\beq
\sigma_W(R)^2 \equiv \int_V \frac{\d\bx_1}{V} \frac{\d\bx_2}{V} \; \DW(\bx_1,\bx_2) .
\label{sigW}
\eeq
However, contrary to the Gaussian case we now have:
\beq
\lag \dL(\bx_1) \dL(\bx_2) \rag \neq \DW(\bx_1,\bx_2) \; , \hspace{0.2cm} \lag \delta_{L,R}^2 \rag \neq \sigma_W(R)^2.
\eeq
The form (\ref{psiNG1}) corresponds to a large density contrast tail of the form $\cP_L(\delta_{L,R}) \sim e^{-|\delta_{L,R}|^{\alpha_N}}$ for the linear density field.

In order to compute the Laplace transform $\psi(y)$ we simply follow the steps of the calculation developed in \cite{paper2}. Thus, we first define a new generating function $\varphi(y)$ through:
\beq
\psi(y) = e^{-\varphi(y \sigma_W^{\alpha_N},\sigma_W)/\sigma_W^{\alpha_N}}
\label{psiNG2}
\eeq
in order to factorize the amplitude of the kernel $\DW$. This yields from eq.(\ref{psiNG1}):
\beq
e^{-\varphi(y,\sigma_W)/\sigma_W^{\alpha_N}} = \cN \int [\d\dL(\bx)] \; e^{- S_W /\sigma_W^{\alpha_N}}
\label{phiNG1}
\eeq
where the ``action'' $S_W[\dL]$ is given by:
\beq
S_W[\dL] = y \; \dR[\dL] + \sum_{i=1}^{N} \lambda_i \; \sigma_W^{\alpha_N} (\dL . \DW^{-1} . \dL)^{\alpha_i/2} .
\label{SW}
\eeq
The quasi-linear regime corresponds to the limit $\sigma_W \rightarrow 0$ at fixed $y$. Then, we see that all terms with $1 \leq i \leq N-1$ in the action $S_W$ in eq.(\ref{SW}) vanish as $\propto \sigma_W^{\alpha_N-\alpha_i}$ since $\DW^{-1} \propto \sigma_W^{-2}$, see eq.(\ref{sigW}). Therefore, in the limit $\sigma_W \rightarrow 0$ we are left with the action $\Sa$:
\beq
\Sa[\dL] = y \; \dR[\dL] + \frac{\sigma_W^{\alpha}}{\alpha} \; (\dL . \DW^{-1} . \dL)^{\alpha/2}
\label{SNG1}
\eeq
with $\alpha=\alpha_N$ and we normalized $\DW$ such that $\lambda_N=1/\alpha$. The fact that the terms with $1 \leq i \leq N-1$ disappear in the quasi-linear limit expresses the fact that this limit is actually a ``rare-event limit''. Indeed, as seen in \cite{paper2} for the Gaussian case (and this remains valid here) a finite $y$ corresponds to a finite density contrast. Then, in the limit $\sigma_W \rightarrow 0$ these finite density contrasts (even though small) become very rare events. On the other hand, the tails of the functional distribution $\cP[\dL]$ defined by the weight $W[\dL]$ are governed by the highest power in eq.(\ref{Wd1}). This implies that in the quasi-linear limit the only relevant term is the highest power $\alpha_N$. Note that we can expect the validity of the quasi-linear limit to extend to larger values of $\sigma_W$ in the case $N=1$ than for $N>1$. Indeed, in this latter case we can expect the terms $1 \leq i \leq N-1$ to make a significant contribution for $\sigma_W \sim 1$ which is not taken into account at all by the quasi-linear limit.

\subsection{Steepest-descent method}
\label{Steepest-descent method}

The action $\Sa$ is independent of the normalization of the kernel $\DW$. Thus, as in \cite{paper2} we can apply the steepest-descent method in the limit $\sigma_W \rightarrow 0$. Indeed, it is clear than in this limit the path-integral (\ref{phiNG1}) is dominated by the global minimum of the action $\Sa[\dL]$ while the contributions from other points $\dL$ are exponentially damped. Then, the steepest-descent method yields asymptotically exact results in this limit. We briefly describe below the main steps of this steepest-descent method. Since the derivation is very close to the one performed in \cite{paper2} for the Gaussian case we refer the reader to that paper for the details of the calculation. Note that the Gaussian case actually corresponds to $\alpha=2$.

In order to apply the steepest-descent approximation we first need to find the global minimum of the action $\Sa[\dL]$. The condition which expresses that the point $\dL$ is an extremum (or a saddle-point) is:
\beq
\frac{\delta \Sa}{\delta(\dL(\bx))} = 0 \hspace{0.3cm} \mbox{for all} \hspace{0.3cm} \bx ,
\label{extremum}
\eeq
where $\delta/\delta(\dL(\bx))$ is the functional derivative with respect to $\dL$ at the point $\bx$. This constraint also writes:
\beqa
\dL(\bx') & = & \frac{- y}{\sigma_W^{\alpha}} \;  (\dL.\DW^{-1}.\dL)^{1-\alpha/2} \nonumber \\ & & \times  \int \d\bx'' \; \DW(\bx',\bx'') \; \frac{\delta(\dR)}{\delta(\dL(\bx''))} .
\label{dLG1}
\eeqa
Thus, we obtain exactly the same profile for the saddle-point as in the Gaussian case. Indeed, we can see that $\alpha \neq 2$ only modifies a numerical multiplicative factor in the r.h.s. of eq.(\ref{dLG1}) (i.e. independent of $\bx'$ and $\bx''$) which could formally be absorbed into $y$ through $y \rightarrow y \sigma_W^{2-\alpha} (\dL.\DW^{-1}.\dL)^{1-\alpha/2}$. Then, as in \cite{paper2} we obtain a spherically symmetric saddle-point. Note that the existence of a spherical saddle-point could be expected a priori, since the initial conditions are homogeneous and isotropic and we study the density contrast within spherical cells. Therefore, the very problem we investigate is spherically symmetric. However, a priori this spherical saddle-point is not necessarily the global minimum of the action (in full generality it might as well be a maximum). This will need to be checked a posteriori.

The fact that we obtain a spherical saddle-point greatly simplifies the problem since it means that this point $\dL$ is described by the well-known spherical collapse solution of the gravitational dynamics. This reads:
\beq
\left\{ \begin{array}{l}
{\displaystyle \dR = \cF \left[ \delta_{L,R_L} \right] } \\ \\
{\displaystyle R_L^3 = (1+\dR) R^3 }
\end{array} \right.
\label{F1}
\eeq
where the function $\cF[\delta_{L,R_L}]$ is given by the usual spherical collapse solution of the equations of motion (e.g., \cite{Peebles1}, \cite{paper2}). The second equation in (\ref{F1}) merely expresses the conservation of mass and the fact that the matter enclosed within the radius $R$ in the actual non-linear density field actually comes from a Lagrangian comoving radius $R_L$, for a spherical initial condition. The eq.(\ref{F1}) provides the functional $\dR[\dL(\bx)]$ for spherical states $\dL(\bx)$. As shown in \cite{paper2} this is sufficient to derive the spherical saddle-point which obeys the condition (\ref{dLG1}). Using the results of \cite{paper2} with the substitution $y \rightarrow y \sigma_W^{2-\alpha} (\dL.\DW^{-1}.\dL)^{1-\alpha/2}$ we eventually obtain:
\beqa
\delta_{L,R_L} & = & - \; \frac{y}{\sigma_W^{\alpha-2} (\dL.\DW^{-1}.\dL)^{(\alpha-2)/2}} \nonumber \\ & & \times \frac{ \cF' \left[ \delta_{L,R_L} \right] \sigma_W^2(R_L)/\sigma_W^2(R) }{ 1 - \cF' \left[ \delta_{L,R_L} \right] \frac{R^3}{3R_L^2} \delta_{L,R_L} \frac{1}{\sigma_W(R_L)} \; \frac{\d \sigma_W}{\d R}(R_L) }
\label{col4}
\eeqa
together with:
\beq
\delta_{L,R'} = \delta_{L,R_L} \; \frac{\DW(R',R_L)}{\DW(R_L,R_L)} .
\label{col3}
\eeq
These two equations fully define the spherically symmetric saddle-point $\dL(\bx)$. The implicit equation (\ref{col4}) determines $\delta_{L,R_L}$ while the radial profile of this initial state is given by eq.(\ref{col3}). In eq.(\ref{col3}) we introduced the spherically averaged kernel $\DW(R_1,R_2)$ defined by:
\beq
\DW(R_1,R_2) \equiv \int_{V_1} \frac{\d\bx_1}{V_1} \int_{V_2} \frac{\d\bx_2}{V_2} \; \DW(\bx_1,\bx_2) .
\label{DWR1}
\eeq
As in \cite{paper2} these equations can be simplified by introducing the functions $\tau(\dL)$ and $\cG(\tau)$ defined by:
\beq
\tau(\dL) \equiv \frac{- \; \dL \; \sigma_W(R)}{\sigma_W \left[ (1+\cF[\dL])^{1/3} R \right] }
\label{tau2}
\eeq
and:
\beq
\cG(\tau) \equiv \cF\left[\dL(\tau)\right] = \dR
\label{G1}
\eeq
where $\dL(\tau)$ is defined by eq.(\ref{tau2}). Using eq.(\ref{tau2}), we note that for a power-law ``power-spectrum'' $P(k) \propto k^n$ we have $\sigma_W(R) \propto R^{-(n+3)/2}$ so that eq.(\ref{G1}) simplifies to:
\beq
\cG(\tau) = \cF\left[ - \tau \; (1+\cG[\tau])^{-(n+3)/6} \right] .
\label{G3}
\eeq
Note however that in this non-Gaussian case the ``power-spectrum'' $P(k)$ does not obey eq.(\ref{Pk1}). Nevertheless, it still provides a measure of the power associated with the wavenumber $k$.

Finally, the limiting generating function $\varphi(y)$ defined by $\varphi(y) \equiv \varphi(y,\sigma=0)$ is given by the value of the action $\Sa$ at this minimum which yields the implicit system:
\beq
\left\{ \begin{array}{l}
{\displaystyle \tau |\tau|^{\alpha-2} = - y \; \cG'(\tau) } \\ \\
{\displaystyle \varphi(y) = y \; \cG(\tau) + \frac{|\tau|^{\alpha}}{\alpha} }
\end{array} \right.
\label{phiNG2}
\eeq
Of course, this result is similar to the Gaussian case which can be recovered by taking $\alpha=2$. Note that eq.(\ref{phiNG2}) has only been derived for real $y$ (hence $\tau$ is real), so that $|\tau|$ is the absolute value of $\tau$.

Thus, the steepest-descent method allows us to compute the Laplace transform $\psi(y)$ in the quasi-linear regime through eq.(\ref{phiNG2}) which yields $\psi(y)$ as in eq.(\ref{psiNG2}). This will give the pdf $\cP(\dR)$ through eq.(\ref{P1}). However, we first need to check that the spherical saddle-point defined by eq.(\ref{col4}) and eq.(\ref{col3}) is the global minimum of the action $\Sa$. For small positive $y$ this can be rigorously proved as for the Gaussian case, using the fact that the kernel $\DL^{-1}$ is positive definite. We shall not repeat this discussion here (see Sect.3.4 in \cite{paper2}). For negative $y$ this cannot be proved as it may actually happen that the spherical saddle-point is only a local minimum. This also occurs in the Gaussian case for a power-spectrum with $n<0$. This case is discussed in great details in \cite{paper2} where we show that the steepest-descent method remains useful but requires a careful justification. Since the same discussion can be applied to the non-Gaussian model we investigate here we refer the reader to \cite{paper2} for a description of such cases and we shall only give a brief comment on this point in the next section.

\subsection{Geometrical construction}
\label{Geometrical construction}

As for the Gaussian case studied in \cite{paper2} we can give a geometrical construction of the generating function $\varphi(y)$ defined by eq.(\ref{phiNG2}). Indeed, this latter expression can also be written:
\beq
\varphi(y) = \min_{\tau} \left[ y \; \cG(\tau) + \frac{|\tau|^{\alpha}}{\alpha} \right] .
\label{phiNG3}
\eeq
The minimum which appears in this expression merely expresses the fact that by sheer definition of the steepest-descent method the generating function $\varphi(y)$ is governed by the global minimum of the action $\Sa$.

Then, we see that the geometrical construction displayed in Fig.4 in \cite{paper2} still applies to this non-Gaussian case, with the modification that the parabola $h-\tau^2/(2y)$ are replaced by the curves $h-|\tau|^{\alpha}/(\alpha y)$. That is, the minimum point $\tau$ which yields $\varphi(y)$ in eq.(\ref{phiNG3}) is given by the first contact of the curve $\cG(\tau)$ with the curves $h-|\tau|^{\alpha}/(\alpha y)$ of varying height $h$. For $y>0$ we start from below at $h=-\infty$ while for $y<0$ we start from above at $h=+\infty$.

As in the Gaussian case, we find that for some values of $n$ the action $\Sa$ has no global minimum. This can be seen from the geometrical construction shown in Fig.4 or simply from eq.(\ref{phiNG3}). Indeed, as noticed in \cite{paper2} from eq.(\ref{G3}) we have the following asymptotic behaviour for $\cG(\tau)$:
\beq
\mbox{high densities} : \; \tau \rightarrow -\infty , \; \cG \rightarrow \infty : \; \cG \sim (-\tau)^{6/(n+3)} .
\label{tauG1}
\eeq
Using this expression we see that for $y<0$ the term in the brackets in the r.h.s. of eq.(\ref{phiNG3}) is not bounded from below if $\alpha < 6/(3+n)$, since in this case it goes to $-\infty$ for $\tau \rightarrow -\infty$. For the Gaussian case this occurs for $n<0$. However, as explained in \cite{paper2} the steepest-descent method developed in Sect.\ref{Steepest-descent method} remains useful and the pdf $\cP(\dR)$ is still governed by the spherical saddle-point defined by eq.(\ref{col4}) and eq.(\ref{col3}). This can be seen by the following remark. If we consider the quantity $\dR^{1/q}$ rather than $\dR$, where $q$ is an odd integer, we can again apply the steepest-descent method which yields the associated generating function $\varphi_q(y)$ as in eq.(\ref{phiNG3}). However, the function $\cG_q(\tau)$ is now given by $\cG_q(\tau) = \cG(\tau)^{1/q}$. Then, if we choose a sufficiently large value for $q$ the behaviour of the r.h.s. bracket in eq.(\ref{phiNG3}) is dominated by the term $|\tau|^{\alpha}$ at large $|\tau|$ so that the action now exhibits a global minimum. Of course, this point is still the spherical saddle-point we obtained in Sect.\ref{Steepest-descent method} (since the initial conditions $\dL$ which give rise to a fixed value of $\dR$ do not depend on whether we study $\dR$ itself or $\dR^{1/q}$ !). Then the pdf $\cP(\dR)$ could be recovered from $\cP(\dR^{1/q})$ through a mere change of variable. This implies in turn that the pdf $\cP(\dR)$ is still governed by the standard spherical saddle-point, even though the latter is no longer the global minimum of the action. However, as described in \cite{paper2} the treatment of this case requires some additional care and it is actually associated with a break-up of perturbative theory.

\subsection{Calculation of the pdf $\cP(\dR)$}
\label{Calculation of the pdf}

Finally, we describe how to compute the pdf $\cP(\dR)$ itself from the results obtained in Sect.\ref{Steepest-descent method}. We consider the case $n=-1$ and $\alpha=3$ where the action $\Sa$ always shows a global minimum. Then, the Laplace transform $\psi(y)$ is fully defined by the generating function $\varphi(y)$ given by eq.(\ref{phiNG2}).

In order to compute the pdf from eq.(\ref{phiNG2}) we simply use the inverse relation (\ref{P1}) which yields:
\beq
\cP(\dR) = \inta \frac{\d y}{2\pi i \sigma_W^{\alpha}} \; e^{[y \dR - \varphi(y)]/\sigma_W^{\alpha}}  .
\label{PNG1}
\eeq
Then, to perform the integration over $y$ we need the analytic continuation of $\varphi(y)$ over the complex plane. Indeed, note that eq.(\ref{phiNG2}) and eq.(\ref{phiNG3}) were actually derived for real $y$ (hence $\tau$ was real). In particular, we must continue the absolute value $|\tau|$ over the complex plane. Moreover, we need to specify the integration path over $y$ (since $\varphi(y)$ will usually be singular at the origin). First, the integration path starts from the real axis at the saddle-point $(\tau_c,y_c)$ given by:
\beq
\frac{\d\chi}{\d y}(y_c) =0 \hspace{0.2cm} \mbox{with} \hspace{0.2cm} \chi(y) \equiv \dR y - \varphi(y) .
\eeq
From eq.(\ref{phiNG2}) we have:
\beq
\cG(\tau_c) = \dR  \;\; \mbox{since} \;\; \varphi'(y) = \cG(\tau)     .
\label{tauc}
\eeq
Thus, as for the Gaussian case, we see from eq.(\ref{G1}) that the triplet $(\dR,\tau_c,y_c)$ is also the triplet $(\dR,\tau,y)$ we obtained in Sect.\ref{Steepest-descent method} to get $\varphi(y)$. This simply means that $\cP(\dR)$ at the point $\dR$ is governed by the neighbourhood of the saddle-point $\dL(\bx)$ obtained in Sect.\ref{Steepest-descent method}, which obeys $\cF(\delta_{L,R_L})=\dR$. This is actually quite natural. Then, we need to continue the function $\varphi(y)$ over the complex plane from the neighbourhood of the point $y_c$ on the real axis. Thus, if $\tau_c>0$ (i.e. $y_c>0$) we replace $|\tau|$ in eq.(\ref{phiNG2}) by $\tau$, while if $\tau_c<0$ (i.e. $y_c<0$) we replace $|\tau|$ by $(-\tau)$. Then, we have the usual analytic continuation of the power-laws $\tau^{\alpha}$ or $(-\tau)^{\alpha}$. The argument of these power-laws is real positive at the starting point $(\tau_c,y_c)$. This provides the analytic continuation of the generating function $\varphi(y)$. The function $\cG(\tau)$ is also analytic around the real axis because the spherical collapse solution $\cF(\dL)$ can be expressed in terms of analytic functions (e.g., trigonometric or hyperbolic functions for a critical density universe).

Next, the integration path in the complex plane follows the steepest-descent contour which runs through the saddle-point $y_c$. It is given by the constraint $\Im(\chi)=0$ and it is orthogonal to the real axis at the point $y_c$. This path is also symmetric about the real axis which implies that the result for $\cP(\dR)$ is real. Note that in the quasi-linear limit $\sigma_W \rightarrow 0$ the contributions to the integral (\ref{PNG1}) only come from an infinitesimal neighbourhood of the saddle-point $(\tau_c,y_c)$ around the real axis. 

Usually the function $\varphi(y)$ is not regular at the origin since we have $\varphi(y) \simeq -(1-1/\alpha) |y|^{\alpha/(\alpha-1)}$ for small real $y$. Then, the pdf $\cP(\dR)$ is not analytic at $\dR=0$ and the moments of the pdf cannot be recovered from the expansion (\ref{psiseries1}). This is not the case for Gaussian initial conditions where both $\varphi(y)$ and $\cP(\dR)$ are regular at the origin.

\begin{figure}[htb]
\centerline{\epsfxsize=8cm \epsfysize=5.8cm \epsfbox{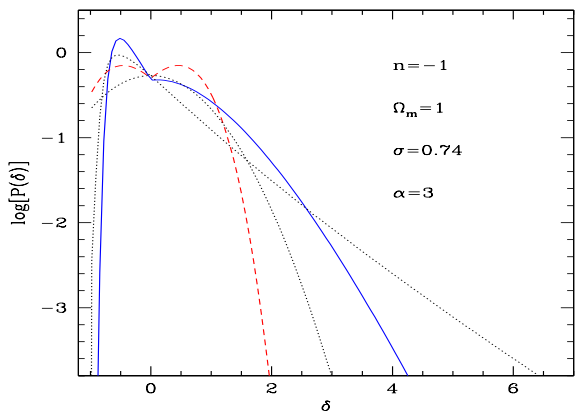}}
\caption{The pdf $\cP(\dR)$ for $n=-1$, $\Om=1$, $\sigma=0.74$ and $\alpha=3$. This corresponds to $\sigma_W=0.65$. The solid line shows the theoretical prediction from (\ref{phiNG2}). The dashed-line displays the linear pdf $\cP_L(\delta_{L,R})$ of the linearly evolved density field. The dotted-lines show the linear and non-linear pdfs obtained for Gaussian initial conditions with the same linear rms density fluctuation $\sigma$ (the non-linear pdf is the asymmetric curve with the extended high-density tail).} 
\label{figPdeltaNG}
\end{figure}

As an example, we show in Fig.\ref{figPdeltaNG} the pdf $\cP(\dR)$ for the case $\alpha=3$ and $n=-1$ (solid line). We also have $\sigma =0.74$ where $\sigma^2 \equiv \lag \delta_{L,R}^2 \rag$ is the rms density fluctuation. Since $\sigma \neq \sigma_W$ we numerically compute $\sigma$ from the second moment of the linear pdf $\cP_L$. Indeed, the steepest-descent method developed in Sect.\ref{Steepest-descent method} also yields the linear pdf $\cP_L(\delta_{L,R})$ of the linearly-evolved density field. In this case we have $\dR=\dL$ and $R_L=R$, hence we simply need to take $\cG(\tau) = -\tau$ in eq.(\ref{phiNG2}) (of course, for $\alpha=2$ this gives back the Gaussian). The non-linear evolution of the density field increases the value of $\cP(\dR)$ at large positive $\dR$ (it actually spreads the linear pdf towards larger value of the density) as in the Gaussian case but the cutoff remains much sharper. It also induces a cutoff at low densities $\dR \geq -1$ which expresses the fact that the actual density $\rho$ is positive.

For comparison, we also display in Fig.\ref{figPdeltaNG} the results we obtain for Gaussian initial conditions with the same linear variance $\sigma$ (dotted lines). These curves were derived in \cite{paper2}. As could be anticipated from eq.(\ref{psiNG1}) the large density contrast cutoff of the linear pdf is much sharper than for the Gaussian case: it goes as $\cP(\delta_{L,R}) \sim e^{-|\delta_{L,R}|^3}$ since here we take $\alpha=3$. Moreover, we can check that for the actual non-linear pdfs this trend remains valid. More precisely, as shown in \cite{paper2} the high-density tail of the non-linear pdf for Gaussian initial conditions is of the form:
\beq
\mbox{Gaussian scenario} : \; \dR \gg 1 : \; \cP(\dR) \sim e^{-\dR^{(n+3)/3}/\sigma^2}
\label{tailG}
\eeq
while for non-Gaussian initial conditions we obtain:
\beq
\mbox{non-Gaussian} : \; \dR \gg 1 : \; \cP(\dR) \sim e^{-\dR^{\alpha(n+3)/6}/\sigma_W^{\alpha}} .
\label{tailNG}
\eeq 
These behaviours are obtained from eq.(\ref{phiNG2}), eq.(\ref{PNG1}) and eq.(\ref{tauc}) which yield:
\beq
\cP(\dR) \sim e^{-|\tau_c|^{\alpha}/(\alpha \sigma_W^{\alpha})}
\eeq
while the asymptotic behaviour of $\tau_c(\dR)$ for large $\dR$ is given by eq.(\ref{tauG1}) and eq.(\ref{tauc}).

As noticed above, the non-Gaussian model (\ref{Wd1}) is not derived from a physical scenario of the primordial universe. It should be viewed as a simple model to describe initial density fluctuations which exhibit a non-Gaussian tail. Then, the analysis performed in the previous sections shows how one may derive in a rigorous manner the gravitational dynamics of density fluctuations in the quasi-linear regime. Moreover, we can expect the form (\ref{tailNG}) of the high-density tail to apply to any model which obeys $\cP(\delta_{L,R}) \sim e^{-|\delta_{L,R}|^3}$ at large densities, whatever the details of the model. In fact, we have shown that in the quasi-linear regime the tail (\ref{tailNG}) (actually the generating function $\varphi(y)$ itself !) is common to all models of the class described by eq.(\ref{Wd1}) with $\alpha_N=\alpha$.

In practice, since observations have not shown any deviations from Gaussianity so far, the parameter $\alpha$ is constrained to be close to 2 (the Gaussian value). Note that this family of models presents the advantage of containing the usual Gaussian scenario as a particular case.

\section{An isocurvature CDM model}
\label{IsoCDM model}

In addition to generalized non-Gaussian models as the one studied in Sect.\ref{Generalization of the Gaussian weight} which are rather ad-hoc tools, there exist some specific non-Gaussian models which arise from a physically motivated description of the primordial universe. These scenarios are of great interest as they could provide a reasonable alternative to the standard simple inflationary model. Moreover, it is prudent not to disregard a priori sensible models of structure formation in view of the rather indirect character of the probes of the early universe which are available to us.

\subsection{Initial conditions}
\label{Initial conditions}

Thus, in this section we consider a specific isocurvature cold dark matter model which was presented in \cite{Peebles3}. This is an inflationary scenario which involves three scalar fields and it gives birth to non-Gaussian isocurvature CDM fluctuations. We refer the reader to \cite{Peebles3} for a description of the physical processes which give rise to this scenario. Then, at the end of inflation (at time $t_i$) the CDM mass distribution is (\cite{Peebles3}):
\beq
\rho_i(\bx) \propto \phi_i(\bx)^2 ,
\label{isorho1}
\eeq
where $\phi_i(\bx)$ is a Gaussian random field with zero mean and power-spectrum $P_{\phi_i}$:
\beq
\lag \phi_i(\bk_1) \phi_i(\bk_2) \rag \equiv P_{\phi_i}(k_1) \delta_D(\bk_1+\bk_2) .
\label{isoP1}
\eeq
On the scales of interest for structure formation we have:
\beq
P_{\phi_i}(k) \propto k^{\np} \; , \; \np \simeq -2.4
\label{isoP2}
\eeq
while below the coherence length $x_c$ (i.e. for $k \gg 1/x_c$) the power-spectrum decreases as $P_{\phi_i}(k) \propto k^s$ with $s<-3$. This coherence length ensures that $\lag \phi_i(\bx)^2 \rag$ is finite, as required by eq.(\ref{isorho1}). In the case of the model described in \cite{Peebles3} we have $x_c \simeq 10$ pc. In the linear regime below the horizon the CDM density perturbations grow as $\dL(\bx,t) \propto D_+(t)$ (we only keep the growing mode) hence we write:
\beq
\dL(\bx,t) = D_+(t) \left( \phi_i(\bx)^2 -1 \right)
\label{isodelta0}
\eeq
where we normalized $\phi_i(\bx)$ by $\lag \phi_i(\bx)^2 \rag = 1$ and the growing mode by $D_+(t_i)=1$. Note that in eq.(\ref{isodelta0}) and in the following we only consider ``small'' comoving scales $x \la 14$ Mpc, where the isocurvature CDM transfer function is close to unity (\cite{Peebles4}). On larger scales the density fluctuations are damped with respect to eq.(\ref{isodelta0}) and we should take into account the decrease of the transfer function. Since we are interested in the statistics of the density field at an arbitrary time $t$ it is convenient to rescale the Gaussian field $\phi_i$. Thus, we define $\phi(\bx) \equiv D_+(t)^{1/2} \phi_i(\bx)$ so that eq.(\ref{isodelta0}) now reads:
\beq
\dL(\bx) = \phi(\bx)^2 - \lag \phi^2 \rag
\label{isodelta1}
\eeq
at the redshift of interest. The power-spectrum $P_{\phi}(k)$ is still of the form (\ref{isoP2}). Note that the relation between the fields $\dL(\bx,t)$ and $\phi(\bx,t)$ depends explicitly on time through the last term in eq.(\ref{isodelta1}) (to simplify notations we usually do not write explicitly the time dependence of the various random fields). Indeed, we must recall here that $\lag \phi^2 \rag$ is not the spatial average of $\phi(\bx)^2$ but the mean over all realizations of the Gaussian random field $\phi$. Therefore, it is independent of the peculiar realization $\phi$ which gives rise to a given linear density field through eq.(\ref{isodelta1}) (even though by ergodicity both averages happen to be equal). We define the two-point correlation of the field $\phi(\bx)$ by:
\beq
\Dp(\bx_1,\bx_2) \equiv \lag \phi(\bx_1) \phi(\bx_2) \rag
\label{Dp1}
\eeq
and the linear variance over the scale $R$ by:
\beq
\sigp^2(R) \equiv \int_V \frac{\d\bx_1}{V} \frac{\d\bx_2}{V} \; \Dp(\bx_1,\bx_2) .
\label{sigp1}
\eeq
Note that eq.(\ref{Dp1}) also implies that we can define the average $\lag \phi^2 \rag$ by:
\beq
\lag \phi^2 \rag \equiv \Dp(\bx,\bx)
\label{Dp0}
\eeq
which does not depend on $\bx$ since the initial conditions are invariant through translations. From eq.(\ref{isodelta1}) and the fact that $\phi(\bx)$ is Gaussian one obtains the variance $\sigma(R)$ of the linearly-evolved density contrast as:
\beq
\sigma^2(R) \equiv \lag \delta_{L,R}^2 \rag = 2 \int_V \frac{\d\bx_1}{V} \frac{\d\bx_2}{V} \; \Dp(\bx_1,\bx_2)^2
\label{isosig1}
\eeq
since the two-point correlation $\DL$ of the linear density field is given by:
\beq
\DL(\bx_1,\bx_2) \equiv \lag \dL(\bx_1) \dL(\bx_2) \rag = 2 \Dp(\bx_1,\bx_2)^2 .
\label{DLDp1}
\eeq
From eq.(\ref{sigp1}) and eq.(\ref{isosig1}) we see that for $-3<\np<-3/2$ we get $\sigma(R) \propto \sigp^2(R)$ and the slope $n$ of the linear power-spectrum $P(k)$ of the density fluctuations is related to $\np$ by: $n=3+2\np$. This implies $\np \simeq -2.4$ as in eq.(\ref{isoP2}) in order to match observational constraints (\cite{Peebles4}). For numerical calculations we shall use:
\beq
\np=-2.4 \;\;\; \mbox{and} \;\;\; n=3+2\np=-1.8
\label{np1}
\eeq

\subsection{Laplace transform}
\label{Laplace transform}

We again define the generating function $\psi(y)$ as in eq.(\ref{psi1}) but the Gaussian average over $\dL(\bx)$ which appeared in eq.(\ref{psi2}) must now be replaced by a Gaussian average over $\phi(\bx)$:
\beq
\psi(y) = \cN \int [\d\phi(\bx)] \; e^{-y \dR[\phi] -\frac{1}{2} \phi . \Dp^{-1} . \phi} .
\label{isopsi1}
\eeq
Thus, our goal is now to estimate the Laplace transform $\psi(y)$. To do so, we shall not introduce a rescaled generating function $\varphi(y)$ as in eq.(\ref{psiNG2}) and we work directly with $\psi(y)$. First, using eq.(\ref{isodelta1}) it is convenient to introduce the linear density field $\dL(\bx)$ into the expression (\ref{isopsi1}) through the Dirac functional $\delta_D[\lag \phi^2 \rag+\dL(\bx) - \phi(\bx)^2]$:
\beqa
\psi(y) & = & \int [\d\phi(\bx)] [\d\dL(\bx)] \; \delta_D[\lag \phi^2 \rag+\dL(\bx) - \phi(\bx)^2] \nonumber \\ & & \times \; e^{-y \dR[\dL] -\frac{1}{2} \phi . \Dp^{-1} . \phi} .
\label{isopsi2}
\eeqa
Here and in the following we do not write the normalization constant of the path-integral (it will be recovered in the final expressions). Then, using the integral representation of the Dirac functional we obtain:
\beqa
\psi(y) & = & \int [\d\phi(\bx)] [\d\dL(\bx)] [\d\lambda(\bx)] \nonumber \\ & & \times \;  e^{-y \dR[\dL] + i \lambda . (\lag \phi^2 \rag+\dL - \phi^2) -\frac{1}{2} \phi . \Dp^{-1} . \phi}
\label{isopsi3}
\eeqa
where $\lambda(\bx)$ is an auxiliary real field. Note that the exponent is actually quadratic over $\phi$ and we can write eq.(\ref{isopsi3}) as:
\beqa
\psi(y) & = & \int [\d\phi(\bx)] [\d\dL(\bx)] [\d\lambda(\bx)] \nonumber \\ & & \times \;  e^{-y \dR[\dL] + i \lambda . (\lag \phi^2 \rag+\dL) -\frac{1}{2} \phi . (\Dp^{-1}+2 i \Lambda) . \phi}
\label{isopsi4}
\eeqa
where we introduced the kernel:
\beq
\Lambda(\bx_1,\bx_2) \equiv \lambda(\bx_1) \delta_D(\bx_1-\bx_2) .
\label{Lambda1}
\eeq
Then, the Gaussian integration over $\phi(\bx)$ is straightforward and it yields:
\beqa
\psi(y) & =  & \int [\d\dL(\bx)] [\d\lambda(\bx)] \left[ \Det(\Dp^{-1}+2 i \Lambda) \right]^{-1/2} \nonumber \\ & & \times \;  e^{-y \dR + i \lambda . (\lag \phi^2 \rag+\dL)} .
\label{isopsi5}
\eeqa
Note indeed that the factor $\lag \phi^2 \rag$ in eq.(\ref{isopsi4}) does not depend on the random field $\phi(\bx)$ over which we integrate, as shown by eq.(\ref{Dp0}). Next, multiplying the r.h.s. of eq.(\ref{isopsi5}) by $(\Det \Dp)^{-1/2}$  (i.e. we change the implicit normalization constant of the path-integral) and using the relation $\ln(\Det M) = \Tr(\ln M)$ we get:
\beqa
\psi(y) & = & \int [\d\dL(\bx)] [\d\lambda(\bx)] \;  e^{ -\frac{1}{2} \Tr \ln(1+2 i \Lambda \Dp) } \nonumber \\ & & \times \;  e^{-y \dR + i \lambda . (\lag \phi^2 \rag+\dL)} .
\label{isopsi6}
\eeqa

At this point, it is instructive to consider the linear regime when the density field is linearly evolved: $\delta(\bx,t) = \dL(\bx,t) \propto D_+(t)$. In this case, we have:
\beq
y \; \dR = y \int \d\bx \; \frac{\theta(x<R)}{V} \; \dL(\bx)
\label{isolin1}
\eeq
where $\theta(x<R)$ is a top-hat with obvious notations. Then, the integration over $\dL(\bx)$ is straightforward and it yields the Dirac functional $\delta_D[ \lambda(\bx) + i y \theta(x<R)/V ]$. The integration over $\lambda(\bx)$ is now obvious and we obtain for the generating function $\psi_L(y)$ which describes the linear density field (again, the subscript ``L'' refers to ``linear''):
\beq
\psi_L(y) \equiv \lag e^{-y \delta_{L,R}} \rag = e^{y \lag \phi^2 \rag - \frac{1}{2} \Tr \ln(1+2 i \Lambda_y \Dp) }
\label{psiL1}
\eeq
where the matrix $\Lambda_y$ is given by:
\beq
\Lambda_y(\bx_1,\bx_2) \equiv - i \; y \; \frac{\theta(x_1<R)}{V} \; \delta_D(\bx_1-\bx_2) .
\label{Lambday1}
\eeq
The normalization of the result derived from the path-integral in eq.(\ref{psiL1}) is obtained from the condition $\psi_L(0)=0$, as implied by the definition of $\psi_L(y)$ (since $\lag 1 \rag =1$). As shown in eq.(\ref{psiseries1}) the generating function $\psi_L(y)$ gives the moments $\lag \delta_{L,R}^q \rag$ of the pdf $\cP_L(\delta_{L,R})$ of the linear density contrast. However, it is usually convenient to introduce the generating function $\Phi_L(y)$ of the cumulants $\lag \delta_{L,R}^q \rag_c$ which is given by (see any textbook on statistical theory):
\beq
\Phi_L(y) \equiv \ln \left[ \psi_L(y) \right] = \sum_{q=1}^{\infty} \frac{(-y)^{q}}{q!} \; \lag \delta_{L,R}^q \rag_c .
\label{PhiL1}
\eeq
Thus, from eq.(\ref{psiL1}) we obtain the exact expression:
\beq
\Phi_L(y) = \lag \phi^2 \rag y - \frac{1}{2} \; \Tr \; \ln(1+2 i \Lambda_y \Dp) .
\label{PhiL2}
\eeq
One can easily check by expanding the logarithm over $y$ that $\Phi_L(y)$ is actually quadratic over $y$ (see also eq.(\ref{T1}) below). Hence $\lag \delta_{L,R} \rag=0$ as it should. Moreover, we can check that we recover eq.(\ref{isosig1}) for the variance of the linear density contrast.

\subsection{Saddle-point contributions}
\label{Saddle-point contributions}

In the non-linear case it is not so easy to compute the expression (\ref{isopsi6}) since the integration over $\dL(\bx)$ no longer yields a Dirac functional. Besides, even for the linear density field the expression (\ref{PhiL2}) is not very convenient since it is not obvious to compute the logarithm. Thus, for practical purposes it only provides the first few cumulants of $\delta_{L,R}$, which can be obtained by expanding the logarithm as a series over $y$ and computing the trace of each power of $\Lambda_y \Dp$. Hence we can try a steepest-descent approximation to the path-integral (\ref{isopsi6}), in the spirit of the calculations presented in \cite{paper2} for the Gaussian case or in Sect.\ref{Generalization of the Gaussian weight} for a simple non-Gaussian model. More precisely, we may try to evaluate the contributions of all possible saddle-points to the path-integral (\ref{isopsi6}).

First, we can look for the saddle-points of the exponent with respect to $\dL(\bx)$. This yields:
\beq
i \lambda(\bx) = y \; \frac{\delta (\dR)}{\delta(\dL(\bx))} .
\label{coliso1}
\eeq
Since the Gaussian random field $\phi(\bx)$ is homogeneous and isotropic we still have a spherical symmetry and we can again look for a spherical saddle-point. Then, taking the mean of eq.(\ref{coliso1}) over a spherical cell $V'$ we get:
\beq
i \lambda_{R'} = y \int_{V'} \frac{\d\bx'}{V'} \; \frac{\delta (\dR)}{\delta(\dL(\bx'))} .
\label{lambda1}
\eeq
From the very definition of derivatives we can also write eq.(\ref{lambda1}) as:
\beqa
i \lambda_{R'} = \lim_{\epsilon \rightarrow 0} \frac{y}{\epsilon} \left\{ \dR \left[ \delta_{L,R''} + \epsilon \Dla(R',R'') \right] - \dR \left[ \delta_{L,R''} \right] \right\}  \nonumber \\
\label{isocol1}
\eeqa
where $R''$ is a dummy variable and we introduced the kernel:
\beqa
\Dla(R',R'') = \int_{V''} \frac{\d\bx''}{V''} \; \frac{\theta(x''<R')}{V'} = \left\{ \begin{array}{l} \frac{1}{V'} \; \mbox{if} \; R'' < R' \\ \frac{1}{V''} \; \mbox{if} \; R'' > R' \end{array} \right. \nonumber \\
\eeqa
We again define the linear density contrast $\delta_{R_L}$ and the Lagrangian mass scale $R_L$ as in eq.(\ref{F1}) and we obtain:
\beq
i \lambda_{R'} = y \; \frac{ \cF' \left[ \delta_{L,R_L} \right] \Dla(R',R_L) }{ 1 - \cF' \left[ \delta_{L,R_L} \right] \frac{R^3}{3R_L^2} \left. \frac{\d \delta_{L,R''}}{\d R''}\right|_{R_L} } .
\label{isocol2}
\eeq
in a fashion similar to the calculation presented in \cite{paper2} for the Gaussian case. The only term in the r.h.s. of eq.(\ref{isocol2}) which depends on $R'$ is the factor $\Dla(R',R_L)$ hence we have:
\beq
\lambda_{R'} \propto \Dla(R',R_L)
\label{isocol3}
\eeq
which yields the profile:
\beq
\lambda(\bx) = \lambda_0 \; \frac{\theta(x<R_L)}{V_L}
\label{lambda2}
\eeq
which also defines the normalization $\lambda_0$. Note that this profile for the field $\lambda(\bx)$ is of the same form as the one obtained for the linear regime which involved the Dirac functional $\delta_D[ \lambda(\bx) + i y \theta(x<R)/V ]$, see the derivation of eq.(\ref{psiL1}). 

Then, for such a state $\lambda(\bx)$ we can estimate the trace which appeared in eq.(\ref{isopsi6}) as follows. Let us note this trace as $T(\lambda_0)$. Then, we have:
\beq
T(\lambda_0) \equiv \Tr \; \ln(1+2 i \Lambda \Dp) = \sum_{q=1}^{\infty} \frac{(-1)^{q+1}}{q} \; (2i)^q \; T_q
\label{Tla0}
\eeq
with:
\beqa
T_q & \equiv & \Tr \; (\Lambda \Dp)^q \nonumber \\ & = & \int \d\bx_1 .. \d\bx_q \; (\Lambda \Dp)(\bx_1,\bx_2) .. (\Lambda \Dp)(\bx_q,\bx_1)
\eeqa
using the definition of the trace: $\Tr M = \int \d\bx M(\bx,\bx)$. Next, from eq.(\ref{Lambda1}) we see that for the spherical state (\ref{lambda2}) the matrix $(\Lambda \Dp)$ is given by:
\beq
(\Lambda \Dp)(\bx_1,\bx_2) = \lambda_0 \; \frac{\theta(x_1<R_L)}{V_L} \; \Dp(\bx_1,\bx_2)
\label{Lambda0}
\eeq
which yields:
\beq
T_q = \lambda_0^q \int_{V_L} \frac{\d\bx_1}{V_L} .. \frac{\d\bx_q}{V_L} \; \Dp(\bx_1,\bx_2) .. \Dp(\bx_q,\bx_1) .
\label{Tq1}
\eeq
The first term $T_1$ (i.e. $q=1$) is simply:
\beq
T_1 = \lambda_0 \int_{V_L} \frac{\d\bx_1}{V_L} \; \Dp(\bx_1,\bx_1) = \lambda_0 \lag \phi^2 \rag
\label{T1}
\eeq
where we used eq.(\ref{Dp1}). Note that it is set by the short-distance behaviour ($x \ll x_c$) of the two-point correlation $\Dp$, since we actually have $x=0$. On the other hand, for $\np < -3/2$ the other terms $T_q$ with $q \geq 2$ are set by the long-distance behaviour of the correlation $\Dp$, i.e. by the scale $R$ one is looking at by studying $\dR$. This is the case for the power-spectra $P_{\phi}$ of cosmological interest, see eq.(\ref{isoP2}). Moreover, for large $q$ the different factors $\Dp$ in eq.(\ref{Tq1}) are almost uncorrelated and it makes sense to use the approximation:
\beq
q \geq 2 : \; T_q \simeq \lambda_0^q \; \left[ \sigp^2(R_L) \right]^q .
\label{Tq2}
\eeq
Besides, for a power-law power-spectrum $P_{\phi}(k)$ as in eq.(\ref{isoP2}) we have (at scales larger than the coherence length $x_c$):
\beq
\Dp(\bx_1,\bx_2) \propto |\bx_1-\bx_2|^{-(\np+3)} .
\label{Dpn}
\eeq
Then, since $\np=-2.4$ is close to $-3$ the correlation $\Dp(\bx_1,\bx_2)$ is almost constant over most of the integration volume so that the approximation (\ref{Tq2}) should be quite reasonable. This yields for the trace $T(\lambda_0)$ defined in eq.(\ref{Tla0}):
\beq
T(\lambda_0) \simeq 2 i \lambda_0 \lag \phi^2 \rag - 2 i \lambda_0 \sigp^2(R_L) + \ln \left[ 1 + 2 i \lambda_0 \sigp^2(R_L) \right] .
\label{Tla01}
\eeq

Next, we could try to consider a saddle-point with respect to both $\dL(\bx)$ and $\lambda(\bx)$. However, one can check that this procedure does not give meaningful results. In fact, as described above in the derivation of the exact eq.(\ref{psiL1}) the generating function $\psi_L(y)$ of the linear density field itself is not governed by a unique saddle-point. Indeed, we had to take into account the contributions from all real fields $\dL(\bx)$ in the path-integral to get the Dirac functional which eventually provides the correct result. Then, in order to keep this structure while using the spherical saddle-points (\ref{lambda2}) it is natural to try to approximate the path-integral (\ref{isopsi6}) by:
\beqa
\psi(y) & \simeq & \int_{-\infty}^{\infty} \frac{\d\delta_{L,R_L}}{2\pi} \int_{-\infty}^{\infty} \d\lambda_0 \; e^{ - \; \frac{1}{2} \; \Tr \; \ln(1+2 i \Lambda \Dp) } \nonumber \\ & & \times \; e^{-y \dR(\delta_{L,R_L}) + i \lambda_0 (\lag \phi^2 \rag +\delta_{L,R_L}) } .
\label{isopsi7}
\eeqa
Here $\dR(\delta_{L,R_L})=\cF(\delta_{L,R_L})$ is given by the spherical collapse solution of the equation of motions, while the matrix $\Lambda \Dp$ is given by eq.(\ref{Lambda0}). Thus, we have replaced the integration over the real fields $\dL(\bx)$ and $\lambda(\bx)$ in eq.(\ref{isopsi6}) by an integration over the real numbers $\delta_{L,R_L}$ and $\lambda_0$. In other words, we have replaced the average over $\dL(\bx)$ and $\lambda(\bx)$ by an average over spherical saddle-points of the form (\ref{lambda2}). In the case of the linear regime when $\delta(\bx) = \dL(\bx)$ we simply have $\dR=\delta_{L,R_L}$ and $R_L=R$. Then we can perform the integration over $\delta_{L,R_L}$ which yields the Dirac function $\delta_D[\lambda_0+iy]$. The integration over $\lambda_0$ is now straightforward and we recover exactly the rigorous result (\ref{psiL1}) (this also justifies the normalization factor $2\pi$ we introduced in eq.(\ref{isopsi7})). Note in addition that the structure of the calculation is also the same (the Dirac functional is simply replaced by a Dirac function). Thus, the expression (\ref{isopsi7}) is actually exact in the linear regime. We show in App.\ref{Reduction to ordinary integrals} how to rigorously derive the expression (\ref{isopsi7}) from eq.(\ref{isopsi6}) in the linear regime. We also explain how it can be derived for the fully non-linear density field in the quasi-linear limit. Thus, eq.(\ref{isopsi7}) is actually exact to leading order in the limit $\sigma \rightarrow 0$. More precisely, for the non-linear density field it yields the exact exponential cutoff of the pdf $\cP(\dR)$ which dominates the dependence on $\sigma$ but the normalization is only correct up to a multiplicative factor of order unity which may depend on $\dR$ but not on $\sigma$. This factor arises from the jacobians $J_a$ and $J_b$ in eq.(\ref{psibas1}). We shall come back to this point below in eq.(\ref{isoP3}).

As explained above, the factor $\Tr \ln(1+2 i \Lambda \Dp)$ is not very convenient for practical calculations hence it is natural to use the approximation (\ref{Tla01}) which was derived for such spherical fields $\lambda(\bx)$. This yields:
\beqa
\psi(y) \simeq \int \frac{\d\dL \d\lambda}{2\pi}  e^{-y \dR + i \lambda \dL + i \lambda \sigp^2(R_L) - \frac{1}{2} \ln(1+2i\lambda \sigp^2(R_L) )} \nonumber \\
\label{isopsi8}
\eeqa
where we removed the subscripts of the integration variables $\dL$ and $\lambda$. Note that in this approximation the term $\lag \phi^2\rag$ has disappeared. In the linear regime, where $\dR=\dL$ and $R_L=R$, we can again perform the integrations over $\dL$ and $\lambda$ which yields:
\beq
\Phi_L(y) \simeq y \; \sigp^2(R) - \frac{1}{2} \; \ln \left( 1+2 y \; \sigp^2(R) \right) .
\label{PhiL3}
\eeq
Then, expanding the logarithm around $y=0$ we obtain the cumulants $\lag \delta_{L,R}^q \rag_c$ from eq.(\ref{PhiL1}). This leads to $\lag \delta_{L,R} \rag_c=0$ and:
\beq
q \geq 2 : \; \lag \delta_{L,R}^q \rag_c \simeq 2^{q-1} \; (q-1)! \; \sigp^{2q}(R) .
\label{isocum1}
\eeq
In particular, we obtain for the variance: $\sigma^2(R) \simeq 2 \sigp^4(R)$, which must be compared with eq.(\ref{sigp1}) and eq.(\ref{isosig1}). Of course, the deviation from the exact result (\ref{isosig1}) is entirely due to the approximation (\ref{Tq2}). For a power-law power-spectrum $P_{\phi}$ as in eq.(\ref{isoP2}) with $\np=-2.4$ we obtain for this deviation $\cD$:
\beq
\cD \equiv \frac{\sigma^2}{2 \sigp^4} = \frac{ \int_V \frac{\d\bx_1}{V} \frac{\d\bx_2}{V} \; \Dp(\bx_1,\bx_2)^2 }{ \left[ \int_V \frac{\d\bx_1}{V} \frac{\d\bx_2}{V} \; \Dp(\bx_1,\bx_2) \right]^2 } \simeq 1.11
\label{isoD1}
\eeq
where we used eq.(\ref{Dpn}) and the result (see \cite{PeebGroth}):
\beq
\int_V \frac{\d\bx_1}{V} \frac{\d\bx_2}{V} \; | \bx_1 - \bx_2 |^{-\gamma} = \frac{72 \; (2R)^{-\gamma}}{(3-\gamma)(4-\gamma)(6-\gamma)} .
\eeq
Thus, we see that the approximation (\ref{Tq2}) is quite reasonable. Note that it becomes exact in the limit $\np \rightarrow -3$. Then, from eq.(\ref{isocum1}) we get for the skewness and the kurtosis:
\beq
D_3 \equiv \frac{\lag \delta_{L,R}^3 \rag_c}{\sigma^3} \simeq 2.83 , \; D_4 \equiv \frac{\lag \delta_{L,R}^4 \rag_c}{\sigma^4} \simeq 12
\eeq
since we have:
\beq
\lag \delta_{L,R}^3 \rag_c \simeq 8 \; \sigp^6 , \; \lag \delta_{L,R}^4 \rag_c \simeq 48 \; \sigp^8  .
\eeq
On the other hand, for the same power-spectrum ($\np=-2.4$) \cite{Peebles4} obtains by a direct numerical calculation (i.e. without the approximation (\ref{Tq2})):
\beq
D_3 \simeq 2.46 , \; D_4 \simeq 9.87
\eeq
and:
\beq
\frac{\lag \delta_{L,R}^3 \rag_c}{8 \; \sigp^6} \simeq 1.02 , \; \frac{\lag \delta_{L,R}^4 \rag_c}{48 \; \sigp^8} \simeq 1.02
\eeq
Thus, we see that the approximation (\ref{Tq2}) is quite satisfactory. In particular, the behaviour of the high-order cumulants $\lag \delta_{L,R}^q \rag_c$ is very well reproduced as soon as $q \ga 3$. Then, we can expect the expression (\ref{isopsi8}) to provide a good approximation to $\psi(y)$, both for the linear and the quasi-linear regimes, since we do not add any approximation in order to describe the non-linear effects encoded in the relations $\delta_{L,R} \leftrightarrow \dR$ and $R_L \leftrightarrow R$.

Going back to eq.(\ref{isopsi8}), we can make the change of variable $\lambda \rightarrow -i z/ \sigp^2(R_L)$ which yields:
\beqa
\psi(y) \simeq \int_{-\infty}^{\infty} \frac{\d\dL}{\sigp^2(R_L)} \inta \frac{\d z}{2\pi i} \; \frac{ e^{-y \dR + z [1+ \dL/ \sigp^2(R_L)]} }{\sqrt{1+2 z}} . \nonumber \\
\label{isopsi9}
\eeqa
Then, after a change of variable and using the integral representation of the Euler Gamma function:
\beq
\frac{1}{\Gamma(\nu)} = \inta \frac{\d z}{2\pi i} \; z^{-\nu} \; e^z
\eeq
we obtain:
\beqa
\psi(y) & \simeq & \int_{-\infty}^{\infty} \frac{\d\dL}{\sqrt{2\pi} \sigp^2(R_L)} \; \theta \left( 1 + \frac{\dL}{\sigp^2(R_L)} > 0 \right) \nonumber \\ & & \times \; e^{-y \dR} \; e^{- \frac{1}{2} [1+ \dL/ \sigp^2(R_L)]} \; \left[ 1 + \frac{\dL}{\sigp^2(R_L)} \right]^{-1/2}
\label{isopsi10}
\eeqa
where the factor $\theta$ is the Heaviside function with obvious notations. Let us recall that in eq.(\ref{isopsi10}) we have: $\dR(\dL) = \cF(\dL)$ and $R_L = R(1+\dR)^{1/3}$.

\subsection{The pdf $\cP_L(\delta_{L,R})$ and $\cP(\dR)$}
\label{isopdf}

In fact, since the generating function is written in an integral form we can directly derive the pdf $\cP(\dR)$ using the inverse Laplace transform (\ref{P1}). Indeed, the integration over $y$ simply gives the Dirac function $\delta_D[\dR-\cF(\dL)]$. After a trivial integration over $\delta$ we obtain:
\beqa
\cP(\dR) & \simeq & \theta \left( 1 + \frac{\dL}{\sigp^2(R_L)} > 0 \right) \; \frac{1}{\sqrt{2\pi} \sigp^2(R_L)} \; \frac{1}{\cF'(\dL)} \nonumber \\ & & \times \; \left[ 1 + \frac{\dL}{\sigp^2(R_L)} \right]^{-1/2} \; e^{- \frac{1}{2} [1+ \dL/ \sigp^2(R_L)]}
\label{isoP3}
\eeqa
where the quantity $\dL$ is given by the condition: $\cF(\dL) = \dR$ and $R_L = R (1+\dR)^{1/3}$. In particular, the pdf $\cP_L(\delta_{L,R})$ of the linearly evolved density field is given by:
\beqa
\cP_L(\delta_{L,R}) & \simeq & \theta \left( 1 + \frac{\delta_{L,R}}{\sigp^2(R)} > 0 \right) \; \frac{1}{\sqrt{2\pi} \sigp^2(R)} \nonumber \\ & & \times \; \left[ 1 + \frac{\delta_{L,R}}{\sigp^2(R)} \right]^{-1/2} \; e^{- \frac{1}{2} [1+ \delta_{L,R}/ \sigp^2(R)]} .
\label{isoPL1}
\eeqa
It is obtained from eq.(\ref{isoP3}) by setting $\cF(\dL)=\dL$ and $R_L=R$. For Gaussian initial conditions the pdf $\cP_L(\delta_{L,R})$ exhibits a specific scaling over the variable $\nu = \delta_{L,R}/\sigma(R)$, which contains all the time and scale dependence of the pdf. As described above, within the approximation (\ref{Tq2}) we have $\sigma^2 \simeq 2 \sigp^4$ hence we get:
\beq
\nu \equiv \frac{\dL}{\sigma(R)} = \frac{\dL}{\sqrt{2} \sigp^2(R)} .
\label{nu1}
\eeq
Then, we see that the linear pdf (\ref{isoPL1}) shows a scaling over $\nu$ as in the Gaussian case:
\beq
\cP_L(\delta_{L,R}) \; \d\delta_{L,R} = \cP_L^{(\nu)}(\nu) \; \d\nu ,
\label{scalnu}
\eeq
but the pdf $\cP_L^{(\nu)}$ is now given by:
\beq
\cP_L^{(\nu)}(\nu) \equiv \frac{\theta(1+\sqrt{2}\nu>0)}{\sqrt{\pi}} \; \frac{1}{\sqrt{1+\sqrt{2}\nu}} \; e^{-\frac{1}{2} (1+\sqrt{2}\nu)} .
\label{Pnu}
\eeq
As noticed in \cite{paper2}, in the Gaussian case the non-linear high-density tail is well described by a simple spherical model. In fact, this model gives the exact exponential dependence of the pdf on the variance $\sigma$, at leading order in the limit $\sigma \rightarrow 0$, as was also shown in \cite{Val1}. Let us recall here how to build this simple model. It is based on the approximation:
\beq
\int_{\dR}^{\infty} \d\delta \; (1+\delta) \cP(\delta) \simeq \int_{\delta_{L,R_L}}^{\infty} \d\dL \; \cP_L(\dL)
\label{spher1}
\eeq
which merely states that the fraction of matter within spherical cells of radius $R$ with a non-linear density contrast larger than $\dR$ is approximately equal to the fraction of matter which was originally enclosed within spherical cells of radius $R_L$ with a linear density contrast larger than $\delta_{L,R_L}$. Here $R_L$ and $\delta_{L,R_L}$ are related to the non-linear variables $R$ and $\dR$ by the usual relation (\ref{F1}), as in eq.(\ref{isoP3}). Note that this is similar to the usual Press-Schechter prescription (\cite{PS}), without the factor 2. Then, substituting the scaling (\ref{scalnu}) into eq.(\ref{spher1}) and differentiating with respect to $\dR$ we obtain:
\beq
\cP_s(\dR) = \frac{1}{1+\dR} \; \frac{\d\nu}{\d\dR} \; \cP_L^{(\nu)}(\nu) , \;\; \mbox{with} \;\; \nu = \frac{\delta_{L,R_L}}{\sigma(R_L)} .
\label{Ps1}
\eeq
Here the subscript ``s'' refers to the ``spherical'' model. In order to recover the variables used in the Gaussian case (\cite{paper2}) it is convenient to introduce the variable $\tau$ given by:
\beq
\tau \equiv - \frac{\delta_{L,R_L} \sigma(R)}{\sigma(R_L)} = - \nu \; \sigma(R) ,
\label{taunu}
\eeq
which removes the dependence on the amplitude of the rms linear density fluctuation, and to define the function $\cG(\tau)$ by:
\beq
\cG(\tau) \equiv \cF[\delta_{L,R_L}] = \dR
\label{Gtau1}
\eeq
which obeys again eq.(\ref{G3}). Then, the relation (\ref{Ps1}) writes:
\beq
\cP_s(\dR) = \frac{1}{1+\dR} \; \frac{1}{\sigma(R)} \; \frac{1}{|\cG'(\tau)|} \;  \cP_L^{(\nu)}(\nu) 
\eeq
which yields:
\beqa
\lefteqn{ \cP_s(\dR) = \theta \left( 1 + \frac{\dL}{\sigp^2(R_L)} > 0 \right) \; \frac{1}{\sqrt{2\pi} \sigp^2(R_L)} \; \frac{1}{1+\dR} } \nonumber \\ & & \times \; \frac{1}{|\cG'(\tau)|} \; \left[ 1 + \frac{\dL}{\sigp^2(R_L)} \right]^{-1/2} \; e^{- \frac{1}{2} [1+ \dL/ \sigp^2(R_L)]} .
\label{isoP4}
\eeqa
We can see that we recover the expression (\ref{isoP3}), except for the multiplicative factor where $\cF'(\delta_{L,R_L})$ has been replaced by $(1+\dR) |\cG'(\tau)|$. Thus, as in the Gaussian case, we find that the simple spherical model (\ref{spher1}) yields the exact exponential $\sigma$-dependent term of the pdf. However, the multiplicative factor may not be exact. Here, we note that the multiplicative factor which appears in eq.(\ref{isoP3}) is not exact either, as stated above, below eq.(\ref{isopsi7}), see also App.\ref{Reduction to ordinary integrals}. In the Gaussian case, we checked by a comparison with numerical simulations (e.g., \cite{Val1}, \cite{paper2}) that the spherical model (\ref{spher1}) provides good results up to $\sigma \simeq 1$. We can expect a similar accuracy in the non-Gaussian case studied here. Therefore, the expression (\ref{isoP4}) should give a good description of the non-linear pdf. This means that the multiplicative factor which appears in eq.(\ref{isoP4}) gives a reasonable approximation of the term induced by the Jacobians which arise in the non-linear case, see eq.(\ref{psibas1}).

\begin{figure}[htb]
\centerline{\epsfxsize=8cm \epsfysize=5.8cm \epsfbox{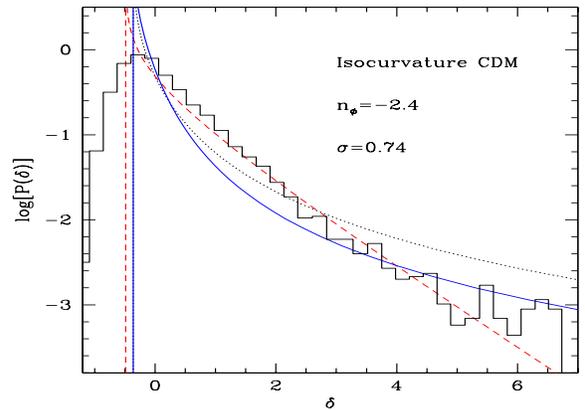}}
\caption{The pdf $\cP(\dR)$ for the isocurvature CDM scenario, with $\np=-2.4$ and $\sigma=0.74$. This corresponds to $\sigp=0.70$. The solid line shows the theoretical prediction from (\ref{isoP4}) for $\cP(\dR)$. This corresponds to the ``spherical'' model. The dotted line is given by eq.(\ref{isoP4}) where no attempt is made to take into account the Jacobians which arise in the non-linear case. The solid line (spherical model) should give better results. The dashed-line displays the pdf $\cP_L(\delta_{L,R})$ of the linearly evolved density field, from eq.(\ref{isoPL1}). The histogram shows the results of numerical simulations by \cite{Rob1} for the linear pdf $\cP_L(\delta_{L,R})$.} 
\label{figPdeltaiso}
\end{figure}

We show in Fig.\ref{figPdeltaiso} our results for $\np=-2.4$ and $\sigma=0.74$. The dashed-line shows the pdf $\cP_L(\delta_{L,R})$ of the linearly evolved density field, from eq.(\ref{isoPL1}). As could be seen from eq.(\ref{isoPL1}) it is strongly non-Gaussian, whatever the value of $\sigp$. In particular, for large overdensities $\delta_{L,R}$ the pdf exhibits a simple exponential cutoff (multiplied by a power $\delta_{L,R}^{-1/2}$) which is much smoother than a Gaussian cutoff. Hence the number of extreme events (i.e. $\delta_{L,R} \gg \sigma$) is much larger than for the Gaussian case, as can be checked by comparison with the lower dotted curve in Fig.\ref{figPdeltaNG}. We can see that for $\delta_{L,R} \geq 0$ our predictions agree very well with the results of numerical simulations from \cite{Rob1} for the linear density field (shown by the histogram). Note that for the theoretical curve we used the exact value of $\sigp$ obtained from eq.(\ref{isoD1}) in order to get a meaningful comparison. Hence the variance of the approximate pdf (\ref{isoPL1}) underestimates the exact result by a factor $\cD \simeq 1.11$. However, we can check that the agreement is quite good for $\delta_{L,R} \geq 0$ so that for practical purposes one can directly use eq.(\ref{isoPL1}). For instance, \cite{Rob1} actually use the linear pdf to apply the usual Press-Schechter recipe (\cite{PS}) in order to estimate the mass function of just-collapsed objects. 

On the other hand, we note that our estimate (\ref{isoPL1}) fails for $\delta_{L,R} \la 0$. In particular, we see that the pdf vanishes for $\delta_{L,R} < - \sigp^2(R)$, that is the exact lower bound $\delta_{L,R} \geq -\lag \phi^2 \rag$ has been replaced by $\delta_{L,R} \geq - \sigp^2(R)$. This is due to the approximation (\ref{Tq2}). Indeed, the exact linear pdf obtained from eq.(\ref{P1}) and eq.(\ref{psiL1}) is:
\beq
\cP_L(\delta_{L,R}) = \inta \frac{\d y}{2\pi i} \; e^{y (\delta_{L,R}+\lag \phi^2 \rag) - \frac{1}{2} \Tr \; \ln(1+2 i \Lambda_y \Dp) } .
\label{isoPL2}
\eeq
This expression clearly shows that the linear density contrast obeys the lower bound $\delta_{L,R} \geq -\lag \phi^2 \rag$. Indeed, for $\delta_{L,R} < -\lag \phi^2 \rag$ we can push the integration path over $y$ to the right (i.e. Re$(y) \rightarrow +\infty$) so that the integral vanishes (we have $\Lambda_y \propto - i y$). Note that this behaviour is rather different from the Gaussian case where the linear density contrast can take negative values of arbitrarily large amplitude. In fact, the lower bound for $\delta_{L,R}$ can be directly seen from eq.(\ref{isodelta1}). Note that this bound is directly related to the time $t_i$ at the end of inflation where eq.(\ref{isorho1}) holds. Indeed, from eq.(\ref{isodelta0}) we see that $\lag \phi^2 \rag = D_+(t)$ where the growing mode has been normalized by $D_+(t_i)=1$. This minimum value of $\delta_{L,R}$ is also overestimated by the numerical simulations which only start at a redshift $z_i \sim 39$. Nevertheless, it is clear that the approximate bound $\delta_{L,R} \geq -\sigp^2$ gives the value below which underdensities become very rare (i.e. the pdf should decline for $\delta_{L,R} \la -\sigp^2$). Besides, for practical purposes one is often mainly interested in the behaviour of overdensities $\dR>0$ so that the expressions (\ref{isoP3}) and (\ref{isoPL1}) should remain useful. 

As noticed in eq.(\ref{scalnu}), the linear pdf given by eq.(\ref{isoPL1}) exhibits a scaling over the one variable $\nu$ defined in eq.(\ref{nu1}), as in the usual Gaussian case. This scaling property was actually used in order to obtain $\cP_L(\delta_{L,R})$ from the numerical simulations. However, this property is not exact and it fails for underdensities since as noticed above the linear density contrast also satisfies the lower bound $-\lag \phi^2 \rag$ which does not scale as $\sigma$ (e.g., contrary to $\sigma$ it is independent of scale).

Next, we show the theoretical prediction $\cP(\dR)$ from eq.(\ref{isoP3}) (dotted line) and from eq.(\ref{isoP4}) (solid line) from the spherical model (\ref{spher1}) for the actual non-linear density field. In Fig.\ref{figPdeltaiso} we used the approximation:
\beq
\cF(\delta) \simeq \left( 1 - \frac{\delta}{1.5} \right)^{-1.5} - 1
\label{Fcolap1}
\eeq
which has been shown to provide a very good fit to the exact spherical collapse solution for all values of $\Om$ and $\Ol$, see Fig.2 in \cite{Ber4}. This means that the dependence on $\Om$ and $\Ol$ is negligible (in the quasi-linear regime where we do not consider virialized objects) and the pdf only depends on the rms fluctuation $\sigma$ (and the slope $\np$ of the power-spectrum). Unfortunately, there are no available results from numerical simulations to compare with our prediction. However, in view of the reasonable accuracy of our prediction for the linear pdf $\cP_L(\delta_{L,R})$ and the good results obtained for the Gaussian case (see \cite{paper2}) we can expect the expression (\ref{isoP4}) to provide a good estimate of the exact pdf $\cP(\dR)$. It should perform better than eq.(\ref{isoP3}) where no attempt was made to take into account the Jacobians which arise in the non-linear case, see App.\ref{Reduction to ordinary integrals}. Of course, we recover the usual features of the non-linear evolution which increases the high-density tail of the pdf. Note again the much larger number of high density events with respect to the Gaussian case with the same variance (see Fig.\ref{figPdeltaNG}).

\section{Standard slow-roll inflation}
\label{Standard slow-roll inflation}

\subsection{Initial conditions}
\label{Initial eps}

The two classes of models we investigated in Sect.\ref{Generalization of the Gaussian weight} and Sect.\ref{IsoCDM model} are strongly non-Gaussian (except for $\alpha \simeq 2$ in the first case) and they apply to ``non-standard'' scenarios for the generation of primordial density fluctuations. For instance, the second model is based on a multifield inflationary scenario. By contrast, in standard slow-roll inflation the primordial density perturbations should be very close to Gaussian. Nevertheless, they may still show some small deviations from Gaussianity. In particular, one is led to consider models where the perturbed primordial gravitational potential $\Phi(\bx)$ is given by:
\beq
\Phi(\bx) = \phi(\bx) + \epsilon \left( \phi(\bx)^2 - \lag \phi^2 \rag \right)
\label{Phi1}
\eeq
where $\phi(\bx)$ is a Gaussian random field with zero mean and power-spectrum $P_{\phi}(k)$, as in eq.(\ref{isoP1}). Such models arise in standard slow-roll inflation if we keep track of the perturbations of the inflaton up to the second-order (e.g., \cite{Gan1}, \cite{Falk1}). In fact, eq.(\ref{Phi1}) can be seen as the first two terms of a Taylor expansion so that it applies to a large class of slightly non-Gaussian models. Thus, the parameter $\epsilon$ is taken to be small so that the primordial density fluctuations are close to Gaussian. Note that $\epsilon$ has the dimensions of $\phi^{-1}$. We shall define below what we mean by ``$\epsilon$ being small'' (see eq.(\ref{Phi2})). Then, our goal here is to obtain the pdf $\cP(\dR)$ of the density fluctuations in the quasi-linear regime up to first order in $\epsilon$ (to go beyond this order we should first take into account the possible higher-order terms in eq.(\ref{Phi1})). Note that for $\epsilon=0$ we must recover the case of Gaussian primordial density fluctuations studied in \cite{paper2}. The gravitational potential $\Phi(\bx)$ is related to the primordial linear density fluctuations $\delta_L(\bx)$ by the Poisson equation (in comoving coordinates):
\beq
\Delta \Phi = \frac{4 \pi \cG \rhob}{a} \; \delta_L(\bx)
\label{Poisson}
\eeq
where $a(t)$ is the scale-factor and $\rhob$ the mean comoving density. Taking the Fourier transform of eq.(\ref{Poisson}) one can eventually write the linear density field $\dL(\bx)$ at the time of interest as:
\beq
\dL(\bx) = \int \d\bx' \; W(\bx,\bx') \left[ \phi(\bx') + \epsilon \left( \phi(\bx')^2 - \lag \phi^2 \rag \right) \right]
\label{Weps1}
\eeq
where the kernel $W(\bx,\bx')$ is given by:
\beqa
W(\bx,\bx') = - \frac{2}{3} \Om^{-1} H_0^{-2} D_+(t) \int \frac{\d\bk}{(2\pi)^3} \; e^{i \bk . (\bx-\bx')} k^2 T(k) \nonumber \\
\label{Weps2}
\eeqa
Here we introduced the Hubble constant $H_0$ and the density parameter $\Om$ today, while $D_+(t)$ is the usual linear growing mode at the time of interest, normalized by $D_+(t_0)=1$ today. The function $T(k)$ is simply the adiabatic CDM transfer function (normalized to unity for $k \rightarrow 0$). Indeed, contrary to the isocurvature model described in Sect.\ref{IsoCDM model} we need to take into account the deviations of $T(k)$ from unity on the scales of interest for large-scale structure formation. The factor $k^2$ arises from the Laplacian in the l.h.s. in eq.(\ref{Poisson}). Note that the kernel $W(\bx,\bx')$ is homogeneous, isotropic and symmetric since it only depends on $|\bx-\bx'|$. Hence for any real fields $f_1$ and $f_2$ we have:
\beqa
\lefteqn{ (f_1 . W . f_2) = (f_2 . W . f_1) = \int \d\bx_1\d\bx_2 W(\bx_1,\bx_2) f_1(x_1) f_2(x_2) .} \nonumber \\
\label{scal1}
\eeqa
Thus, eq.(\ref{Weps1}) defines the initial conditions of our system.

We can note that the pdf $\cP_L(\delta_{L,R})$ of the linearly evolved density field was already studied in \cite{Mat1} using path-integral methods like those described in Sect.\ref{Generating function} below. However, they did not investigate the effects of the non-linear dynamics. On the other hand, \cite{Ver1} studied the observational tests which may constrain such deviations from Gaussian initial conditions.

\subsection{Generating function}
\label{Generating function}

In order to derive the pdf $\cP(\dR)$ we again introduce the Laplace transform $\psi(y)$ as in eq.(\ref{psi1}), which yields again eq.(\ref{isopsi1}). Next, we introduce the linear density field through the Dirac functional $\delta_D[\dL(\bx) - \int \d\bx' W(\bx,\bx') \{ \phi(\bx') + \epsilon ( \phi(\bx')^2 - \lag \phi^2 \rag ) \} ]$. This Dirac functional can again be expressed through an auxiliary real field $\lambda(\bx)$ so that the analog of eq.(\ref{isopsi3}) now reads:
\beqa
\psi(y) & = & \int [\d\phi(\bx)] [\d\dL(\bx)] [\d\lambda(\bx)] \; e^{-y \dR[\dL] -\frac{1}{2} \phi . \Dp^{-1} . \phi} \nonumber \\ & & \times \;  e^{i \lambda . (\dL - W . [ \phi + \epsilon ( \phi^2 - \lag \phi^2 \rag ) ] ) }
\label{eps1}
\eeqa
where we defined the inverse $\Dp^{-1}$ of the two-point correlation $\Dp$ of the Gaussian random field $\phi$, which is again given by eq.(\ref{Dp1}). The mean $\lag \phi^2 \rag$ is again given by eq.(\ref{Dp0}) and it does not depend on the field $\phi(\bx)$ over which we integrate. As in Sect.\ref{IsoCDM model} we do not write explicitly the normalization factor of the path-integrals. Next, we introduce the kernel $\Lambda_W(\bx_1,\bx_2)$ defined by:
\beq
\Lambda_W(\bx_1,\bx_2) \equiv \delta_D(\bx_1-\bx_2) \int \d\bx \; \lambda(\bx) W(\bx,\bx_1) 
\label{LaW1}
\eeq
so that eq.(\ref{eps1}) reads:
\beqa
\psi(y) & = & \int [\d\phi(\bx)] [\d\dL(\bx)] [\d\lambda(\bx)] \; e^{-y \dR[\dL] -\frac{1}{2} \phi . \Dp^{-1} . \phi } \nonumber \\ & & \times \;  e^{i \lambda . \dL - i \lambda . W . \phi - i \epsilon \phi . \Lambda_W . \phi + i \epsilon \lag \phi^2 \rag (\lambda . W . 1)} .
\label{eps2}
\eeqa
As explained in Sect.\ref{Initial eps} we only consider the first-order term in $\epsilon$. Therefore, we expand the exponential in eq.(\ref{eps2}) up to first-order in $\epsilon$. This yields:
\beqa
\psi(y) & = & \int [\d\phi(\bx)] [\d\dL(\bx)] [\d\lambda(\bx)] \; e^{-y \dR - i \lambda . W . \phi -\frac{1}{2} \phi . \Dp^{-1} . \phi}  \nonumber \\ & & \times \; e^{i \lambda . \dL} \; \left[ 1 + i \epsilon \lag \phi^2 \rag (\lambda . W . 1) - i \epsilon (\phi . \Lambda_W . \phi) \right] .
\label{eps3}
\eeqa
This path-integral is Gaussian over the random field $\phi$. Thus, we first make the change of variable $\phi = \phi' - i \Dp . W . \lambda$ so that eq.(\ref{eps3}) writes:
\beqa
\lefteqn{ \psi(y) = \int [\d\phi'(\bx)] [\d\dL(\bx)] [\d\lambda(\bx)] \; e^{-y \dR + i \lambda . \dL  -\frac{1}{2} \lambda . W . \Dp . W . \lambda} } \nonumber \\ & & \times \; e^{-\frac{1}{2} \phi' . \Dp^{-1} . \phi'} \; \biggl [ 1 + i \epsilon \lag \phi^2 \rag (\lambda . W . 1) - i \epsilon (\phi' . \Lambda_W . \phi')  \nonumber \\ & &  - 2 \epsilon (\phi' . \Lambda_W . \Dp . W . \lambda) + i \epsilon (\lambda . W . \Dp . \Lambda_W . \Dp . W . \lambda) \biggl ] .
\label{eps4}
\eeqa
Here we used the fact that all kernels $W$, $\Dp$ and $\Lambda_W$ are symmetric. Then, using Wick's theorem we can perform the Gaussian integration over $\phi'(\bx)$ which yields:
\beqa
\psi(y) & = & \int [\d\dL(\bx)] [\d\lambda(\bx)] \; e^{-y \dR + i \lambda . \dL  -\frac{1}{2} \lambda . W . \Dp . W . \lambda}  \nonumber \\ & & \times \biggl [ 1 + i \epsilon (\lambda . W . \Dp . \Lambda_W . \Dp . W . \lambda) \biggl ]
\label{eps5}
\eeqa
where we used $\Dp(\bx,\bx)=\lag \phi^2 \rag$. Next, we define the two-point correlation $\DLo(\bx_1,\bx_2)$ of the linear density field in the case $\epsilon=0$ by:
\beqa
\DLo(\bx_1,\bx_2) \equiv \lag \dL(\bx_1) \dL(\bx_2) \rag_0 = \lag (W.\phi)(\bx_1) (W.\phi)(\bx_2) \rag \nonumber \\
\label{DLeps1}
\eeqa
where we used eq.(\ref{Weps1}). Here the subscript ``0'' refers to ``$\epsilon=0$''. The relation (\ref{DLeps1}) also writes:
\beq
\DLo = W.\Dp.W
\label{DLeps2}
\eeq
The path-integral (\ref{eps5}) is Gaussian over the field $\lambda(\bx)$. Thus, making the change of variable $\lambda = \lambda'+i \DLo^{-1} . \dL$ and integrating over $\lambda'(\bx)$ using Wick's theorem we finally get:
\beqa
\psi(y) & = & \int [\d\dL(\bx)] \; e^{-y \dR[\dL] -\frac{1}{2} \dL.\DLo^{-1}.\dL} \nonumber \\ & & \times \; \left[ 1 + \epsilon \; (\Dp^{-1}.W^{-1}.\dL).(W^{-1}.\dL)^2 \right] 
\label{eps6}
\eeqa
where we used $\int \d\bx \; \dL(\bx)=0$. Here we introduced the short-hand notation $(W^{-1}.\dL)^2$ for the vector:
\beq
(W^{-1}.\dL)^2(\bx) \equiv \left( \int \d\bx' \; W^{-1}(\bx,\bx') \dL(\bx') \right)^2 .
\eeq
Note that for $\epsilon=0$ we recover the expression (\ref{psi2}) derived for Gaussian initial conditions. This was to be expected since for $\epsilon=0$ the linear density field is actually Gaussian. Indeed, in this case we have $\dL=W.\phi$ and $\phi$ is a Gaussian random field. Of course, the procedure we described above can be extended up to any order in $\epsilon$ (provided we know the initial conditions up to the required order). Indeed, by expanding the exponential which appears in eq.(\ref{eps1}) up to the needed order in $\epsilon$ the integrations over the fields $\phi(\bx)$ and $\lambda(\bx)$ are Gaussian and can be easily performed using Wick's theorem. Then, one eventually obtains an expression of the form (\ref{eps6}), where the term in brackets is an expansion over $\epsilon$ up to the required order. Note that this is similar to the standard Edgeworth expansion where one directly expands the Laplace transform $\psi(y)$ around the Gaussian value $\psi_0(y)=e^{y^2 \sigma^2/2}$ (see any textbook on probability theory). Therefore, the expansion over $\epsilon$ does not necessarily converge: it may only provide an asymptotic series (e.g., \cite{Cram1}).

\subsection{Steepest-descent method}
\label{eps Steepest-descent method}

We now need to evaluate the path-integral (\ref{eps6}). As for the Gaussian case studied in \cite{paper2} or the non-Gaussian generalization described in Sect.\ref{Generalization of the Gaussian weight} we can use a steepest-descent method for the quasi-linear regime. We refer the reader to \cite{paper2} for a detailed presentation of this method. First, we define the rescaled generating function $\psib(y)$ by:
\beq
\psib(y) \equiv \psi \left(y/\sigma_0(R)^2\right)
\label{psib1}
\eeq
where $\sigma_0(R)$ is the variance of the linear density field for $\epsilon=0$:
\beq
\sigma_0(R)^2 \equiv \lag \delta_{L,R}^2 \rag_0 = \int_V \frac{\d\bx_1}{V} \frac{\d\bx_2}{V} \; \DLo(\bx_1,\bx_2) .
\label{sig0eps}
\eeq
Using eq.(\ref{eps6}) we get:
\beqa
\psib(y) & = & \int [\d\dL(\bx)] \; e^{-S[\dL]/\sigma_0(R)^2} \nonumber \\ & & \times \; \left[ 1 + \epsilon \; (\Dp^{-1}.W^{-1}.\dL).(W^{-1}.\dL)^2 \right] 
\label{eps7}
\eeqa
where we introduced the ``action'' $S[\dL]$ given by:
\beq
S[\dL] \equiv y \; \dR[\dL] + \frac{\sigma_0(R)^2}{2} \; \dL.\DLo^{-1}.\dL
\label{Seps1}
\eeq
In the quasi-linear limit $\sigma_0 \rightarrow 0$ the path-integral (\ref{eps7}) is governed by the minimum of the action $S$. As shown in \cite{paper2} this spherically symmetric saddle-point is given by:
\beqa
\dL(\bx) & = & \delta_{L,R_L} \int_{V_L} \frac{\d\bx'}{V_L} \; \frac{\DLo(\bx,\bx')}{\sigma_0(R_L)^2} \nonumber \\ & = & \frac{\delta_{L,R_L}}{\sigma_0(R_L)^2} \; (W.\Dp.W) . \left( \frac{\theta(x'<R_L)}{V_L} \right)
\label{coleps1}
\eeqa
where $\bx'$ is a dummy variable. The variable $\delta_{L,R_L}$ is given by the implicit equation:
\beq
\delta_{L,R_L} = - y \; \frac{ \cF' \left[ \delta_{L,R_L} \right] \sigma_0^2(R_L)/\sigma_0^2(R) }{ 1 - \cF' \left[ \delta_{L,R_L} \right] \frac{R^3}{3R_L^2} \delta_{L,R_L} \frac{1}{\sigma_0(R_L)} \; \frac{\d \sigma_0}{\d R}(R_L) } .
\label{coleps2}
\eeq
Here the function $\cF(\delta_{L,R_L})$ is the usual spherical collapse solution while $R_L$ is the Lagrangian mass scale, see eq.(\ref{F1}). These results were also used in Sect.\ref{Steepest-descent method} where we investigated a simple generalization of the Gaussian weight. Next, we introduce the functions $\tau(\dL)$ and $\cG(\tau)$ by:
\beq
\tau(\dL) \equiv \frac{- \; \dL \; \sigma_0(R)}{\sigma_0 \left[ (1+\cF[\dL])^{1/3} R \right] }
\label{epstau1}
\eeq
and:
\beq
\cG(\tau) \equiv \cF\left[\dL(\tau)\right] = \dR
\label{epsG1}
\eeq
as in eq.(\ref{tau2}) and eq.(\ref{G1}). Then, we define the function $\varphi_0(y) \equiv \min S[\dL]$ as the value of the action $S$ at this spherically symmetric saddle-point (i.e. the minimum of the action). As shown in \cite{paper2} it is given by the implicit system:
 \beq
\left\{ \begin{array}{l}
{\displaystyle \tau = - y \; \cG'(\tau) } \\ \\
{\displaystyle \varphi_0(y) = y \; \cG(\tau) + \frac{\tau^2}{2} }
\end{array} \right.
\label{epsphi1}
\eeq
in a fashion similar to eq.(\ref{phiNG2}). Then, at leading order in $\sigma_0$ we can write the path-integral (\ref{eps7}) in the quasi-linear limit as:
\beq
\psib(y) = e^{-\varphi_0(y)/\sigma_0(R)^2} \; \left[ 1 - \epsilon \; \frac{\tau^3}{\sigma_0(R)^3} \; \frac{J_3(R_L)}{\sigma_0(R_L)^3} \right]
\label{eps8}
\eeq
where we defined:
\beq
J_3(R_L) \equiv \left( W . \frac{\theta(x_1<R_L)}{V_L} \right) . \left( \Dp . W . \frac{\theta(x_2<R_L)}{V_L} \right)^2 .
\label{J31}
\eeq
Here $\bx_1$ and $\bx_2$ are dummy variables. The normalization of $\psib(y)$ in eq.(\ref{eps8}) is set by the constraint $\psib(0) = \lag 1 \rag =1$. Next, we define the dimensionless quantity $\tilde{\epsilon}$ by:
\beq
\tilde{\epsilon}(R_L) \equiv \epsilon \; \frac{J_3(R_L)}{\sigma_0(R_L)^3} .
\label{teps1}
\eeq
It is convenient to express the quantities $\sigma_0$ and $J_3$ in terms of the power-spectrum $P_{\phi}(k)$ of the Gaussian random field $\phi(\bx)$. This yields:
\beq
\sigma_0(R_L)^2 \equiv (2\pi)^6 \int \d\bk \; P_{\phi}(k) W(k)^2 F(k R_L)^2
\label{sig0eps2}
\eeq
and:
\beqa
\lefteqn{ J_3(R_L) \equiv (2\pi)^9 \int \d\bk \d\bk' \; P_{\phi}(k) P_{\phi}(k') W(k) W(k') } \nonumber \\ & & \times \; W(|\bk+\bk'|) F(k R_L) F(k' R_L) F(|\bk+\bk'| R_L) .
\label{J32}
\eeqa
Here $W(\bk)=W(|\bk|)$ is the Fourier transform of the window function $W(\bx)$ defined in eq.(\ref{Weps2}) (i.e. it is related to the kernel $W$ by $W(\bx_1,\bx_2)=W(\bx_2-\bx_1)$) while $F(kR)$ is the Fourier transform of the top-hat of radius $R$:
\beq
F(kR) \equiv \int_V \frac{\d\bx}{V} \; e^{i \bk . \bx} = 3 \; \frac{\sin(kR)-(kR)\cos(kR)}{(kR)^3} .
\label{Ftophat1}
\eeq
Note that in eq.(\ref{Ftophat1}) we did not introduce the factor $(2\pi)^{-3}$ used in the definition (\ref{Four1}) of the Fourier transform, in order to obtain the usual top-hat window $F(kR)$. As in the previous sections we approximate the power-spectrum $P_0(k)$ of the linear density fluctuations (when $\epsilon=0$) by a power-law. Since we have $P_{\phi}(k) \propto k^{-3}$ (i.e. a scale-invariant primordial power-spectrum, with small and large scale cutoffs) for the standard inflationary scenarios we investigate here this means that $W(k)^2 \propto k^{n+3}$ on the scales of interest. Indeed, the power-spectrum $P_0(k)$ is related to $P_{\phi}(k)$ by:
\beq
P_0(k) = (2\pi)^6 W(k)^2 P_{\phi}(k) ,
\label{epsPk1}
\eeq
as can be seen for instance from eq.(\ref{sig0eps2}). Then, from eq.(\ref{sig0eps2}) and eq.(\ref{J32}) we get:
\beq
\sigma_0(R_L)^2 \propto R_L^{-(n+3)} , \;\;\; J_3(R_L) \propto R_L^{-3(n+3)/2}
\label{sigRL1}
\eeq
which implies that the quantity $\tilde{\epsilon}(R_L)$ defined in eq.(\ref{teps1}) does not depend on $R_L$, within this approximation. Moreover, from eq.(\ref{sig0eps2}) and eq.(\ref{J32}) we see that we have the order of magnitude estimate:
\beq
\tilde{\epsilon} \sim \epsilon \left( \int \d\bk \; P_{\phi}(k) \right)^{1/2} = \epsilon \; \sqrt{ \lag \phi^2 \rag } .
\label{teps2}
\eeq
Going back to the definition (\ref{Phi1}) of the initial conditions we see that we can write the primordial perturbations of the gravitational potential as:
\beq
\Phi(\bx) = \phi(\bx) + {\cal O}(1) \; \tilde{\epsilon} \; \frac{\phi(\bx)^2 - \lag \phi^2 \rag}{\sqrt{ \lag \phi^2 \rag }}
\label{Phi2}
\eeq
where ${\cal O}(1)$ stands for a numerical factor of order unity. It is clear from this expression that for a scale-invariant primordial power-spectrum the initial conditions are close to Gaussian if $\tilde{\epsilon} \ll 1$. Hence $\tilde{\epsilon}$ is the relevant dimensionless parameter which describes the amplitude of the deviations from Gaussianity and the usual inflationary scenarios lead to $\tilde{\epsilon} \ll 1$.

Here it is interesting to consider the linear regime, that is the statistics of the linear density field. Then, we have $\cG(\tau)=-\tau$ and $\tau=y=-\delta_{L,R} = -\dR$, so that $\varphi_0(y) = -y^2/2$. This yields from eq.(\ref{eps8}):
\beq
\psi_L(y) = e^{y^2 \sigma_0^2/2} \; \left[ 1 - \tilde{\epsilon} \; y^3 \sigma_0^3 \right]
\label{epspsiL1}
\eeq
where the subscript ``L'' refers to the ``linearly evolved'' density field. Then, from the expansion (\ref{psiseries1}) we obtain for the third-order moment of the linear density field $\lag \delta_{L,R}^3 \rag$ and for the skewness $D_3$:
\beq
\lag \delta_{L,R}^2 \rag = \sigma_0^2 , \;\; \lag \delta_{L,R}^3 \rag = 6 \tilde{\epsilon} \sigma_0^3 , \;\; D_3 \equiv \frac{\lag \delta_{L,R}^3 \rag}{\lag \delta_{L,R}^2 \rag^{3/2}} = 6 \tilde{\epsilon}
\label{epsD3}
\eeq
at first-order in $\tilde{\epsilon}$. Thus, we see that the parameter $\tilde{\epsilon}$ is directly related to the skewness of the linear density field on the scale of interest, due to the slight non-Gaussianity of the primordial density field.

\subsection{The pdf of the density contrast}
\label{The pdf of the density contrast}

From the rescaled Laplace transform $\psib(y)$ obtained in eq.(\ref{eps8}) we can derive the pdf $\cP(\dR)$, using the inverse Laplace transform (\ref{P1}). This yields:
\beq
\cP(\dR) = \inta \frac{\d y}{2\pi i \sigma_0^2(R)} \; e^{[y \dR - \varphi_0(y)]/\sigma_0^2} \;  \left[ 1 - \tilde{\epsilon} \; \frac{\tau^3}{\sigma_0^3} \right]
\label{epsP1}
\eeq
where the function $\tau(y)$ is given by eq.(\ref{epsphi1}). The pdf $\cP_L(\delta_{L,R})$ of the linearly evolved density field is obtained from eq.(\ref{epsP1}) by using $\varphi_0(y)=-y^2/2$ and $\tau(y)=y$ which gives:
\beq
\cP_L(\delta_{L,R}) = \inta \frac{\d y}{2\pi i \sigma_0^2} \; e^{[y \delta_{L,R} + y^2/2]/\sigma_0^2} \;  \left[ 1 - \tilde{\epsilon} \; \frac{y^3}{\sigma_0^3} \right] .
\label{epsPL1}
\eeq
Then we can perform the Gaussian integration over $y$ in eq.(\ref{epsPL1}) which yields (see \cite{Grad1}, \S 3.462.4):
\beq
\cP_L(\delta_{L,R}) = \frac{e^{-\delta_{L,R}^2/(2\sigma_0^2)}}{\sqrt{2\pi}\sigma_0(R)} \; \left[ 1 + \tilde{\epsilon} \; 2^{-3/2} \; H_3 \left( \frac{\delta_{L,R}}{\sqrt{2} \sigma_0} \right) \right]
\label{epsPL2}
\eeq
where $H_3(x)$ is the Hermite polynomial of order 3 defined by:
\beq
H_3(x) \equiv 8 x^3 - 12 x .
\label{H3}
\eeq

\begin{figure}[htb]
\centerline{\epsfxsize=8cm \epsfysize=5.8cm \epsfbox{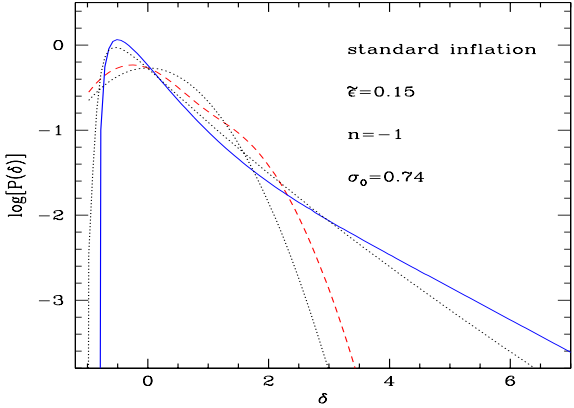}}
\caption{The pdf $\cP(\dR)$ for $n=-1$ and $\sigma_0=0.74$, for slightly non-Gaussian initial conditions: $\tilde{\epsilon}=0.15$. The solid line shows the prediction of eq.(\ref{epsP1}) for the non-linear pdf $\cP(\dR)$ while the dashed-curve shows the pdf $\cP_L(\delta_{L,R})$ of the linearly evolved density field from eq.(\ref{epsPL2}). The dotted lines display the results obtained with $\tilde{\epsilon}=0$, i.e. for Gaussian initial conditions.} 
\label{figPdeltaeps}
\end{figure}

Finally, we display in Fig.\ref{figPdeltaeps} our results for slightly non-Gaussian initial conditions: $\tilde{\epsilon}=0.15$. In order to compare the pdf with the case of Gaussian initial conditions we also show the results obtained for $\tilde{\epsilon}=0$ (dotted curves). Let us recall that eq.(\ref{epsPL1}) for the pdf of the linearly evolved density field is exact to first-order in $\tilde{\epsilon}$. In agreement with eq.(\ref{epsD3}) we can check that $\tilde{\epsilon}>0$ leads to a small positive skewness for the linear density field. In particular, the high-density tail of the pdf is slightly enhanced. We can see that this feature remains valid for the non-linear pdf $\cP(\dR)$. However, as gravitational clustering proceeds the deviations from Gaussianity become dominated by the non-linear dynamics over the slight primordial non-Gaussianity. More precisely, we can obtain the low-order moments of the density field through the expansion (\ref{psiseries1}). At the lowest order in $y$ we have $\tau=y + {\cal O}(y^2)$ and $\varphi_0(y) = -y^2/2! + S_3 y^3/3! + {\cal O}(y^4)$ where the parameter $S_3$ is given by the standard result (e.g., \cite{Ber2}):
\beq
S_3 = \frac{34}{7} - (n+3) .
\eeq
Then, we obtain from eq.(\ref{eps8}) the expansion:
\beq
\psi(y) = 1 + \frac{1}{2!} y^2 \sigma_0^2 - \frac{1}{3!} S_3 y^3 \sigma_0^4 - \tilde{\epsilon} y^3 \sigma_0^3 + {\cal O}(y^4) .
\eeq
This yields for the lowest order moments:
\beq
\lag \dR^2 \rag = \sigma_0^2 \; , \; \lag \dR^3 \rag = S_3 \; \sigma_0^4 + 6 \; \tilde{\epsilon} \; \sigma_0^3
\eeq
and for the skewness $D_3$:
\beq
D_3 \equiv \frac{\lag \dR^3 \rag}{\lag \dR^2 \rag^{3/2}} = 6 \; \tilde{\epsilon} + S_3 \; \sigma_0 .
\eeq
This relation clearly shows that the relative importance of the primordial deviation from Gaussianity is greater at earlier times before gravitational clustering builds up, as could be expected. Note that in Fig.\ref{figPdeltaeps} we used the generating function $\varphi_0(y)$ defined by the branch of eq.(\ref{epsphi1}) which is regular at the origin $y=0$. Since here we considered the case $n=-1$ this function actually shows a branch cut along the real negative axis for $y<y_s$ (with $y_s<0$). Then, for large density contrasts one needs to take into account the second branch of $\varphi_0(y)$ defined over the range $y_s<y<0$ which is singular at the origin. However, as shown in Fig.6 in \cite{paper2} this is irrelevant for the case displayed in Fig.\ref{figPdeltaeps} for $\dR <7$. We refer the reader to \cite{paper2} for a detailed discussion of this point. Of course, for the non-Gaussian model we study here one simply needs to follow the procedure outlined in \cite{paper2} for Gaussian initial conditions and to add to the relevant expressions the term in the brackets in eq.(\ref{epsP1}) which describes the effects due to the small primordial deviation from Gaussianity (to first order in $\tilde{\epsilon}$).

\section{Conclusion}

Thus, using a non-perturbative method developed in a previous work (\cite{paper2}), we have described in details how to obtain the pdf $\cP(\dR)$ of the density contrast within spherical cells in the quasi-linear regime for three specific non-Gaussian models.

The first case is a straightforward generalization of the Gaussian scenario and it can be seen as a phenomenological description of a density field where the tails of the linear pdf are of the form $\cP(\dL) \sim e^{-|\dL|^{\alpha}}$, where $\alpha$ is no longer required to be equal to two (as in the Gaussian case). Then, we have shown that the derivation of the pdf presented for the Gaussian case can be directly extended to this model. This provides again exact results for the pdf in the quasi-linear limit. 

The second scenario is a physically motivated model of isocurvature cold dark matter presented in \cite{Peebles3}. It arises from an inflationary scenario with three scalar fields. This case is slightly more difficult as one needs to adapt the method to this specific model. Moreover, in order to get simple analytical results one must introduce a simple approximation (which is not related to the gravitational dynamics but to the non-Gaussian properties of the initial conditions). However, we have shown that even with this approximation we get good results for the linear pdf $\cP_L(\delta_{L,R})$ for $\delta_{L,R} \ga 0$ by comparison with numerical simulations. For lack of data we could not check our prediction for the non-linear pdf $\cP(\dR)$ but we can expect a good agreement of the same accuracy.

Finally, the third scenario corresponds to the small primordial deviations from Gaussianity which arise in standard slow-roll inflation. We obtained exact results for the pdf of the density field in the quasi-linear limit, to first-order over the primordial deviations from Gaussianity.

This study shows that our approach is powerful enough to be applied to a large variety of initial conditions. Note that our predictions for the linear pdf $\cP_L(\delta_{L,R})$ for such non-Gaussian scenarios may be used to estimate the mass function of just-collapsed objects, using a straightforward extension of the Press-Schechter recipe. Note that we shall discuss the Press-Schechter prescription in the light of the formalism developed in \cite{paper2} and used in this work in a companion article (\cite{paper4}).

To conclude, we note that our approach being non-perturbative it can in principle be applied to the non-linear regime. Indeed, it does not rely either on the hydrodynamical description. We shall present a study of this non-linear regime in a future work, see \cite{paper4}, for Gaussian initial conditions. However, it is clear that this can be extended to the non-Gaussian models described here.

\appendix

\section{Isocurvature scenario. Reduction to ordinary integrals}
\label{Reduction to ordinary integrals}

In this appendix we show how to derive the expression (\ref{isopsi7}) from eq.(\ref{isopsi6}). We first consider the linear regime where eq.(\ref{isopsi7}) is exact. The path-integral (\ref{isopsi6}) involves the functional measure $[\d\dL(\bx)]$. The latter can be defined from a discretization procedure (i.e. a spatial grid for the coordinate $\bx$ with an infinitesimal spacing) but an equivalent formulation (e.g., \cite{Zinn1}) is to expand the function $\dL(\bx)$ on a complete set of real orthonormal functions $f_q$ (in the Hilbert space ${\cal L}^2$):
\beq
\dL(\bx) = \sum_{q=0}^{\infty} a_q \; f_q(\bx) .
\label{dLfq1}
\eeq
The vectors $f_q$ obey the relation:
\beq
\int \frac{\d\bx}{V} \; f_q(\bx) f_r(\bx) = \delta_{q,r}
\label{fqfr1}
\eeq
which expresses the fact that they form an orthonormal basis. Here $\delta_{q,r}$ is the usual Kronecker symbol. We introduced a factor $1/V$ in the scalar product (\ref{fqfr1}) so that the functions $f_q$ are dimensionless, but this is not essential (here $V$ is a constant). Then, the functional measure $[\d\dL(\bx)]$ can be defined as (e.g., \cite{Zinn1}):
\beq
[\d\dL(\bx)] \equiv \cN \prod_{q=0}^{\infty} \d a_q
\label{meas1}
\eeq
where $\cN$ is a normalization constant. In order to get a discrete basis in eq.(\ref{dLfq1}) we restricted the density field to a large finite volume ${\cal V}$. However, this is not essential: we choose ${\cal V}$ to be much larger than any relevant length scale and we can eventually take the limit ${\cal V} \rightarrow \infty$. Next, we can choose for the first basis vector $f_0$ the top-hat of radius $R$:
\beq
f_0(\bx) \equiv \theta(x<R)
\label{f01}
\eeq
where $\theta(x<R)$ is the usual top-hat with obvious notations. Then, from eq.(\ref{dLfq1}) and eq.(\ref{fqfr1}) we have:
\beq
a_0 = \int \frac{\d\bx}{V} \; f_0(\bx) \dL(\bx) = \delta_{L,R} .
\label{a01}
\eeq
We can expand the auxiliary field $\lambda(\bx)$ on the same basis:
\beq
\lambda(\bx) = \sum_{q=0}^{\infty} b_q \; f_q(\bx)
\label{lambdafq1}
\eeq
so that the path-integral (\ref{isopsi6}) now writes:
\beqa
\psi_L(y) & = & \int_{-\infty}^{\infty} \prod_{q=0}^{\infty} \d a_q \prod_{r=0}^{\infty} \d b_r \; e^{-y a_0 + i V \sum_{q=0}^{\infty} a_q b_q} \nonumber \\ & & \times \; e^{i \lag \phi^2 \rag (\lambda . 1) -\frac{1}{2} \Tr \ln(1+2 i \Lambda \Dp) }
\label{psiLbas1}
\eeqa
where the subscript ``L'' refers to the ``linearly evolved'' density field (so that $\dR=\delta_{L,R}=a_0$) and we did not write the normalization constant of the integrals. Then, the integration over the variables $a_q$ with $q \geq 1$ yields the Dirac functions $\delta_D(b_q V)$. The integration over $b_r$ is now straightforward for $r \geq 1$ and we obtain:
\beqa
\psi_L(y) & = & \int_{-\infty}^{\infty} \d a_0 \d b_0 \; e^{-y a_0 + i V a_0 b_0 + i \lag \phi^2 \rag b_0 V} \nonumber \\ & & \times \; e^{ - \frac{1}{2} \Tr \ln(1+2 i \Lambda \Dp) }
\label{psiLbas2}
\eeqa
where the matrix $\Lambda$ is obtained from $\lambda(\bx) = b_0 \theta(x<R)$. Then, using $a_0 = \delta_{L,R}$ from eq.(\ref{a01}) and making the change of variable $\lambda_0 = b_0 V$ we obtain eq.(\ref{isopsi7}) in the linear regime (i.e. $\delta_{L,R_L} = \delta_{L,R} = \dR$ and $R_L=R$).

Now, we investigate how the previous derivation is modified when we study the non-linear density field $\delta(\bx)$. We can again define the functional measure $[\d\dL(\bx)]$ (and $[\d\lambda(\bx)]$) by eq.(\ref{meas1}). However, we now introduce the new variable $\delta_{L,R_L}$, which also defines $\dR \equiv \cF(\delta_{L,R_L})$ and $R_L \equiv R (1+\dR)^{1/3}$. Then, we define the functions $f_q'(\bx)$ by:
\beq
f_q'(\bx) \equiv \sqrt{\frac{V}{V_L}} \; f_q \left(\frac{R}{R_L} \; \bx \right)
\label{fq1}
\eeq
so that we still have the orthonormalization property:
\beq
\int \frac{\d\bx}{V} \; f_q'(\bx) f_r'(\bx) = \delta_{q,r} .
\label{fqfr2}
\eeq
Then, we write the linear density field $\dL(\bx)$ as:
\beq
\dL(\bx) = \sum_{q=0}^{\infty} a_q' \; f_q'(\bx)
\label{dLfq2}
\eeq
which also implies:
\beq
a_0' = \sqrt{\frac{V_L}{V}} \; \delta_{L,R_L} .
\label{a02}
\eeq
This yields:
\beq
\dL(\bx) = \delta_{L,R_L} \theta(x<R_L) + \sum_{q=1}^{\infty} a_q' \; f_q'(\bx)
\label{dLfq3}
\eeq
while we write the auxiliary field $\lambda(\bx)$ as:
\beq
\lambda(\bx) = \sum_{q=0}^{\infty} b_q' \; f_q'(\bx) .
\label{lambdafq2}
\eeq
Then, the path-integral (\ref{isopsi6}) now writes:
\beqa
\psi(y) & = & \int_{-\infty}^{\infty} \d\delta_{L,R_L} \prod_{q=1}^{\infty} \d a_q' \prod_{r=0}^{\infty} \d b_r' \; J_a J_b \; e^{-y \dR} \nonumber \\ & & \times \; e^{i V \sum_{q=0}^{\infty} a_q' b_q'+ i \lag \phi^2 \rag (\lambda . 1) -\frac{1}{2} \Tr \ln(1+2 i \Lambda \Dp) }
\label{psibas1}
\eeqa
where we made the changes of variables $\{a_q\}_{q \geq 0} \rightarrow \{\delta_{L,R_L} , \{a_q'\}_{q \geq 1}\}$ and $\{b_q\}_{q \geq 0} \rightarrow \{b_q'\}_{q \geq 0}$. In eq.(\ref{psibas1}) the jacobians $J_a$ and $J_b$ of these transformations are given by the determinants:
\beq
J_a \equiv \left| \Det \left[ \frac{\pl a_i}{\pl \delta_{L,R_L}} , \frac{\pl a_i}{\pl a_j'} \right] \right| \; , \; J_b \equiv \left| \Det \left[ \frac{\pl b_i}{\pl b_j'} \right] \right| .
\label{jacob1}
\eeq
These jacobians do not depend on the amplitude $\sigma(R)$ of the density fluctuations. On the other hand, the quasi-linear regime corresponds to the limit $\sigma \rightarrow 0$ for a fixed finite $y$ (which leads to finite $\delta_{L,R_L}$ and $\dR$, as shown in the Gaussian case in \cite{paper2}). In this limit, the generating function $\psi(y)$ is governed by the exponent in eq.(\ref{psibas1}) which depends on $\sigma$ and sets the cutoffs of order $\sigma$ for the density contrast $\dR$. Then, we can neglect the Jacobians in eq.(\ref{psibas1}): they do not contribute to $\ln(\psi)$ at leading order for $\sigma \rightarrow 0$ (however, they give a $\sigma$-independent prefactor which appears in $\psi$, see the discussion below). Moreover, if we make the approximation $\dR[\dL] \simeq \cF(\delta_{L,R_L})$ we can again integrate over $a_q'$ for $q' \geq 1$, which gives the Dirac functions $\delta_D(b_q' V)$, and we eventually recover eq.(\ref{isopsi7}).

Let us now discuss the approximation $\dR[\dL] \simeq \cF(\delta_{L,R_L})$ used above. The point is that in the limit $\sigma \rightarrow 0$ the path-integral (\ref{isopsi6}) and the ordinary integrals (\ref{psibas1}) are dominated by the saddle-points of the exponent. As discussed in Sect.\ref{Saddle-point contributions}, because of the spherical symmetry of the physics we investigate there exist some spherical saddle-points, of the form (\ref{lambda2}). Then, if there are no other saddle-points (or if some other saddle-points exist but they yield a lower contribution to integral, which will vanish in the limit $\sigma \rightarrow 0$) the path-integral is given at leading-order by its value for these spherical saddle-points. Note that the case of Gaussian initial conditions studied in \cite{paper2} exhibits the same behaviour in a simpler manner. Indeed, in this case there exists only one spherical saddle-point for both the linear and non-linear generating functions $\psi_L(y)$ and $\psi(y)$. Then, in order to keep only the leading order for $\sigma \rightarrow 0$ one defines the rescaled generating function $\varphi(y)$ by:
\beq
\psi(y) = e^{-\varphi(y \sigma^2)/\sigma^2}
\label{psiGauss}
\eeq
as in eq.(\ref{psiNG2}), which can be expressed through the path-integral:
\beq
e^{-\varphi(y)/\sigma^2} = \int [\d\dL(\bx)] \; e^{- S[\dL]/\sigma^2}
\label{SGauss1}
\eeq
in a fashion similar to eq.(\ref{phiNG1}). Next, in the quasi-linear limit $\sigma \rightarrow 0$ the path-integral in the r.h.s. in eq.(\ref{SGauss1}) is dominated by one saddle-point and taking the logarithm of both sides in eq.(\ref{SGauss1}) one obtains in this limit:
\beq
\varphi(y) = \min_{\dL(\bx)} S[\dL]
\label{SGauss2}
\eeq
as described in \cite{paper2} (see also Sect.\ref{Steepest-descent method}). Going back to $\psi(y)$ this means that in the quasi-linear limit the generating function $\psi(y)$ is given by the maximum of the exponent which appears in the relevant path-integral. This corresponds to the procedure we described above for the isocurvature CDM scenario in order to get eq.(\ref{isopsi7}). However, the difference is that in this non-Gaussian case there exists an infinite number of spherical saddle-points with respect to $\dL(\bx)$. Therefore, at leading order the generating function $\psi(y)$ is not given by a unique contribution but rather by the sum of all contributions due to these saddle-points which are parameterized by the variables $\delta_{L,R_L}$ and $\lambda_0$.

Finally, we note that the minimum of the action $S[\dL]$, or the value of the exponent in eq.(\ref{psibas1}), only provides the leading order contribution to $\ln(\psi)$ in the limit $\sigma \rightarrow 0$. This means that it yields the exact exponential $\sigma$-dependent term of the pdf $\cP(\dR)$ which describes the cutoffs of the pdf for density contrasts which are large relative to $\sigma$. In our case, this is the factor $e^{-(1+\dL/\sigp^2)/2}$ which appears in eq.(\ref{isoP3}). However, it does not give the exact multiplicative prefactor of the generating function $\psi$ and of the pdf $\cP(\dR)$. Indeed, the Jacobians $J_a$ and $J_b$ in eq.(\ref{psibas1}) (or the Gaussian integration around the saddle-point in the Gaussian case) contribute to this term. In the case we study here, this means that eq.(\ref{isopsi7}) differs from the exact result by a multiplicative factor which may depend on $\dL$ and $\lambda$. Therefore, the expression (\ref{isoP3}) obtained for the pdf is correct up to a multiplicative factor which may depend on $\dR$ but not on $\sigma$. This point is also discussed in Sect.\ref{isopdf} where we compare eq.(\ref{isoP3}) with the result (\ref{isoP4}) of a simple spherical model. The latter can be seen as an attempt to model this prefactor. Note that for the pdf of the linear density field we do not encounter this difficulty since, as shown above, eq.(\ref{psiLbas2}) is actually exact.

\end{document}